\documentclass[a4paper,fleqn,usenatbib]{mnras}

\usepackage[T1]{fontenc}
\usepackage{ae,aecompl}

\usepackage{graphicx}	
\usepackage{amsmath}	
\usepackage{amssymb}	

\usepackage{multirow}
\usepackage{booktabs}
\usepackage{times}
\usepackage{caption}
\usepackage{subcaption}
\usepackage{{threeparttable}}

\title[Long-term soft and hard X-ray investigation of WR 25]{Long-term soft and hard X-ray investigation of the colliding wind WN+O binary WR 25}

\author[Arora, Pandey \& De Becker]{
Bharti Arora$^{1,2}$\thanks{E-mail: bharti@aries.res.in},
J. C. Pandey$^{1}$\thanks{E-mail: jeewan@aries.res.in}, and 
M. De Becker$^{2}$\thanks{E-mail: Michael.DeBecker@uliege.be}
\\
$^{1}$Aryabhatta Research Institute of Observational Sciences, Nainital$-$263 002, India.\\
$^{2}$Space Sciences, Technologies and Astrophysics Research (STAR) Institute, University of Li\`ege, Quartier Agora, 19c, All\'ee du 6 A\^out, B5c, \\B-4000 Sart Tilman, Belgium\\
}

\date{Accepted XXX. Received YYY; in original form ZZZ}

\pubyear{2018}
\begin{document}
\label{firstpage}
\pagerange{\pageref{firstpage}--\pageref{lastpage}}
\maketitle

\begin{abstract}
We investigated the long-term behaviour in X-rays of the colliding wind binary WR\, 25, using archival data obtained with \textit{Suzaku}, \textit{Swift}, \textit{XMM-Newton}, and \textit{NuSTAR} spanning over $\sim$16 years. Our analysis reveals phase-locked variations repeating consistently over many consecutive orbits, in agreement with an X-ray emission fully explained by thermal emission from the colliding winds in the 208-d orbit. We report on a significant deviation of the X-ray flux with respect to the 1/D trend (expected for adiabatic shocked winds) close to periastron passage. The absence of a drop in post-shock plasma temperature close to periastron suggests this break in trend cannot be explained in terms of reduced pre-shock velocities in this part of the orbit. Finally, \textit{NuSTAR} data reveal a lack of hard X-ray emission (above 10.0 keV) above the background level. Upper limits on a putative non-thermal emission strongly suggest that the sensitivity of present hard X-ray observatories is not sufficient to detect non-thermal emission from massive binaries above 10 keV, unless the wind kinetic power is large enough to significantly feed particle acceleration in the wind-wind interaction. 

\end{abstract}




\section{Introduction}

Early type stars with initial masses higher than 10 M$_{\odot}$, corresponding to the spectral types from early-B to O, lie in the upper left part of the Hertzsprung$-$Russell diagram. These objects evolve over short lifetimes (typically of the order of a few to 10 Myr) and during the advanced stages of their evolution, they appear as Wolf$-$Rayet (WR) stars. Being among the category of massive stars, these hot stars are very luminous and have strong radiatively driven winds. Radiatively driven winds are the winds driven by the transfer of momentum from the photospheric radiation field through photon scattering by strong UV resonance lines. Only a fraction of the ions are directly accelerated by this process, however, other charged particles constituting the stellar wind are dragged along by Coulomb interactions \citep[see e.g.][for a review on this topic]{Puls2008}. These winds are highly supersonic, traveling with terminal velocities ($v_{\infty}$) $\sim$ 1000$-$3000  km s$^{-1}$ and have mass loss rates ($\dot{M}$) in the range of $\dot{M}$ $\sim$ 10$^{-7}$ to 10$^{-4}$ M$_{\odot}$ year$^{-1}$. As a consequence of these huge stellar winds, massive stars undergo a very significant degree of mass loss during their lifetime, which not only affects the evolution of the star itself \citep[e.g.][]{Smith2014} but also has a strong impact on their surroundings \citep[e.g.][]{Fierlinger2016}.

Massive stars are, more generally, found in binary or higher multiplicity systems with other massive stars. The periods of these binaries range from a couple of days up to many years (or hundreds or thousands of years). When two massive stars are bound together by gravity, their winds interact with each other and part of their kinetic energy is converted into radiation that gives rise to a number of observational signatures spanning a wide range of the electromagnetic spectrum. In this paper, we will focus on the most spectacular observational consequence of the colliding winds which is the X-ray emission that is produced due to the formation of hydrodynamic shocks in the wind collision region (WCR). X-rays provide the invaluable diagnostics of shock physics in colliding-wind massive binaries.

WR 25 (HD 93162) is a bright (V $\sim$ 8.03 mag) WR binary system located in the Carina Nebula region. The spectral classification of WR 25 has always been a matter of debate because it displays dilute WN6-7 emission line spectrum \citep{Walborn1985} combined with strong absorption features \citep{Moffat1978}. Later on WR 25 was classified as WN6ha by \citet{Smith1996}. Looking at various spectral features of WR 25 as observed by many authors, \citet{vanderHucht2001} categorized it as WN6h+O4f. More recently, it has been moved to the class of the "hot slash" objects by \citet{Crowther2011} on the basis of its P-Cygni H$\beta$ profile and was given  O2.5If$^{*}/$WN6 spectral type. \citet{Sota2014} detected a visual companion in WR 25 with a separation of 790 mas and V-band magnitude difference of 5.8 mag using the \textit{Hubble Space Telescope}  observations. Therefore, the latest spectral classification of WR 25 is O2.5If$^{*}/$WN6$+$OB, with no detailed information about the spectral and luminosity class of companion star. WR 25 was seen in X-ray energy range for the first time by \citet{Seward1979} using \textit{Einstein} X-ray observations (0.2-4.0 keV) of the Trumpler 16 open cluster and its surroundings. Later observations of \textit{Einstein} of the same region revealed the ratio of X-ray to bolometric luminosity of $\sim 2\times10^{-6}$ for WR 25 is an order of magnitude higher than observed for other WR stars in same region \citep{Seward1982}. In the X-ray survey of WR stars by \citet{Pollock1987} and \citet{Pollock1995}, WR 25 was again found to be the brightest X-ray source. Significant variability in the optical polarization was noticed by \citet{Drissen1992} and it was suggested that these modulations were caused by the binary motion of the stars in WR 25. \citet{Raassen2003} attempted to investigate the X-ray emission of WR 25 and could not notice any variability in the emission over a period of 10 years. But the presence of the Fe XXV emission line in \textit{XMM-Newton} spectrum of WR 25 pointed towards wind collision occurring in the system. \citet{Pollock2006} recognized the variations in the X-ray emission of WR 25 for the first time and suggested that it is a colliding wind binary system with a period of about 4 years. Later in the same year, \citet{Gamen2006} provided the radial velocity solutions for WR 25 and it was found to be a long period ($\sim$ 208 days) and eccentric (e$=$0.5) binary system. Orbital parameters of WR 25 are given in Table\,\ref{orb_par}.

A more detailed X-ray study of WR 25 was performed by \citet{Pandey2014} using \textit{Swift} and \textit{XMM-Newton} observations spanning over $\sim$10 years. Looking at enhanced X-ray luminosity accompanied with the phase locked modulations, they concluded that WR 25 is a colliding wind binary (CWB) where the X-ray emission is significantly coming from the hot plasma heated by the colliding winds. Their analysis also pointed towards the hints for a 1/D (D is the binary separation) variation of the X-ray luminosity, at least at higher energies (i.e. above 2\,keV), while the variations in the softer energy bands were seen to be modulated by absorption effects. In order to investigate these variations, a detailed monitoring of the X-ray emission as a function of the orbital phase must be conducted. Such an observational study is needed to constrain the stellar winds collision properties and test the present theoretical models which provide a description of the physics of colliding winds. Because colliding-wind binaries span a wide range of stellar and orbital parameters, it is important to study as many such sources as possible in detail, and consequently confront present theories to a significantly relevant sample of observational facts. To date, there are only a few massive binaries with longer periods (longer than $\sim$ 100 days) which were observed with a good phase coverage. In this context, we have executed a deep X-ray study of WR 25 using the long-term archival X-ray data from modern X-ray observatories.

\begin{table}
	\caption{Orbital parameters of WR\,25.\label{orb_par}}
	\begin{center}
		\begin{tabular}{l c r}
			\hline
			\hline
			Parameter & Value & Reference  \\
			\hline
			\vspace*{-0.2cm}\\
			Period (d) & 207.85 $\pm$ 0.02 & 1 \\
			$V_o$ (km s$^{-1}$) &-34.6 $\pm$ 0.5 & 1 \\
			\textit{K} (km s$^{-1}$) & 44 $\pm$ 2 & 1 \\
			eccentricity  & 0.50 $\pm$ 0.02 & 1 \\
			\textbf{$\omega$} (degrees) & 215 $\pm$ 3 & 1 \\
			$T_{periastron}$ (HJD) & 2451598 $\pm$ 1 & 1 \\
			$T_{RVmax}$ (HJD) & 2451654 $\pm$ 1 & 1 \\
			$a$ sin\,\textit{i} ($R_\odot$) & 156 $\pm$ 8 & 1 \\
			$M_{pri}$ $sin^{3}$\,\textit{i} ($M_\odot$) & 75 $\pm$ 7 & 2 \\
			$M_{sec}$ $sin^{3}$\,\textit{i} ($M_\odot$) & 27 $\pm$ 3 & 2 \\
			\hline
		\end{tabular}
		\begin{tablenotes}
			\small
			\item \textbf{Notes:} Here, $V_o$ is the radial velocity,  \textit{K} is the radial velocity amplitude, \textbf{$\omega$} is the longitude of periastron, $T_{periastron}$ is the time of periastron passage, $T_{RVmax}$ is the time of maximum radial velocity, $a$ sin\,\textit{i} is the projected semi-major axis, $M_{pri}$ $sin^{3}$\,\textit{i} and $M_{sec}$ $sin^{3}$\,\textit{i} are the minimum masses of primary and secondary binary components, respectively.
			\item \textbf{References:} (1) \citet{Gamen2006}; (2) \citet{Gamen2008}
		\end{tablenotes}
	\end{center}
\end{table}

This paper is organized as follows: Section\,\ref{observ} summarises the observations used and the data reduction methodology. Section\,\ref{spectra} describes the X-ray spectral properties of WR 25. The X-ray light curve analysis is given in Section\,\ref{lc}. Our main results are discussed in Section\,\ref{disc}, and Section\,\ref{concl} gives the conclusions.  

\section{Observations and Data reduction}\label{observ}
We used X-ray observations of WR 25 made by \textit{Suzaku}, \textit{Swift}, \textit{XMM-Newton}, and \textit{NuSTAR} from July 2000 to August 2016 for a total of 226 epochs. A detailed log of these observations is given in Table~\ref{log}. The orbital phase of each observation was derived by using ephemeris HJD$=$2,451,598.0$+$207.85E \citep{Gamen2006}. We have also included the \textit{Swift} and \textit{XMM-Newton} datasets that were studied by \citet{Pandey2014}. For homogeneity and application of latest calibration, these X-ray data were processed again. The data reduction procedure adopted for each X-ray observatory is explained as follows. 

\begin{table}
	\centering
	\caption{Log of Observations of WR 25.}
	\setlength{\tabcolsep}{4pt}
	\begin{tabular}{c c c c c c r}
		\toprule[1.5pt] 
		Obs. ID & Obs. Date & Start time & Exp.$/$ & Source &  Phase  & Offset  \\ 
		&      &  (UT)           &\multicolumn{1}{c}{Eff. Exp.} & Counts & ($\phi$) & \multicolumn{1}{c}{($'$)} \\ 
		&      &				  &\multicolumn{1}{c}{(ksec.)} &    & \\
		\hline\hline 		
		
		\multicolumn{1}{c}{\textit{Suzaku}}\\
		
		100012010  &2005-08-29 &01:48:03	&49.8  & 16771	&0.689	&7.500	\\ 
		100045010  &2006-02-03 &09:59:30	&21.4  & 4944	&0.450	&6.521	\\ 
		402039010  &2007-06-23 &05:54:08	&58.4  & 24523	&0.879	&5.312	\\
		403035010  &2008-06-10 &01:51:36	&35.4  & 8363	&0.576	&5.682	\\ 
		.. 	    &.. 	&..		&..   &..	&..	&..	\\ 
		..  	    &.. 	&..		&..  &..	&..	&..	\\
		\bottomrule[1.5pt]
		\label{log}	
	\end{tabular}
	\begin{tablenotes}
		\item Notes: Source counts are given for \textit{Suzaku} XIS-1, \textit{Swift} XRT, \textit{XMM-Newton} MOS2, and \textit{NuSTAR} FPMB detectors.  A detailed log of observations is available online.
	\end{tablenotes}
\end{table}
 
\subsection{\textit{Suzaku}}
The \textit{Suzaku} satellite observed Carina Nebula for 10 times from August 2005 to July 2013 taking Eta\,Carinae at the centre of the X-ray Imaging Spectrometer (XIS$-$0, 1 \& 3) field of view. These data were reprocessed using the currently available calibration database by running the task \textit{aepipeline} (version 1.1.0) for individual observation ID. The data was reduced in accordance with the standard screening criteria. Grade 0, 2, 3, 4, and 6 events were used. Events recorded during the South Atlantic Anomaly passages, Earth elevation angle below 5$^\circ$ and Earth day-time elevation angle below 20$^\circ$ were discarded. Hot and flickering pixels were also removed. Barycentric correction to the clean event files was applied by using the task \textit{aebarycen}. A circular region of radius 90$''$ centred at source position was used to extract the source products, which is more than the XIS half power diameter (HPD) of 2$'$. However, other circular regions of radius 45$''$ near the source free region were selected as background regions.  Light curves, spectra, and the response files were obtained, from both 3\,$\times$\,3 and 5\,$\times$\,5 modes collectively, using the XSELECT (version 2.4c) package. Background subtracted light curves and spectra from two front illuminated (FI) XIS chips (XIS$-$0 \& 3) were added using the tasks \textit{lcmath} and \textit{addascaspec}, respectively. Finally, the back illuminated (BI; XIS$-$1) and FI XIS spectra were grouped to have minimum 20 counts per energy bin for further spectral analysis.  

\subsection{\textit{Swift}}
WR 25 has been observed frequently by \textit{Swift} X-Ray Telescope (XRT) from December 2007 $-$ June 2009 and August 2014 $-$ August 2016 in photon counting mode. These long-term observations enabled us to investigate the X-ray emission at almost all the phases of this long period binary system. The observations with exposure time less than 1 ks were discarded due to few raw counts (see also Table \ref{log}). The data was processed using \textit{Swift} \textit{xrtpipeline} (version 0.13.2). This produced cleaned and calibrated event files using calibration files available in mid-November, 2016. The extraction of image, light curve, and spectrum for every observation was done using XSELECT (version 2.4c) package by selecting standard event grades of 0$-$12. Source products were extracted from a circular region of 30$''$ radius. An annular background region of 69$''$ inner and 127$''$ outer radius, around the source region, was chosen for the background estimation. The spectra were binned to have minimum 10 counts per energy bin with GRPPHA. The response matrix file (RMF) provided by the \textit{Swift} team (swxpc0to12s6$\textunderscore$20130101v014.rmf) was used. In order to take bad columns into account, we calculated an ancillary response file (ARF) for each dataset individually using the task \textit{xrtmkarf} by considering the associated exposure map.

\subsection{\textit{XMM-Newton}}
\begin{figure}
	\includegraphics[width=\columnwidth,trim={0.0cm 0.0cm 0.0cm -0.4cm}]{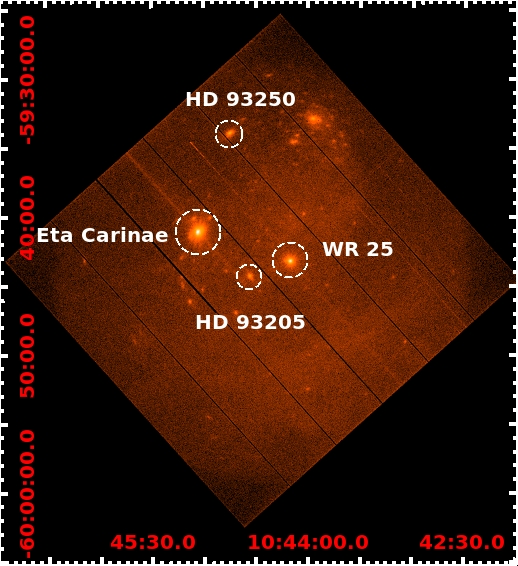}
	\caption{\textit{XMM-Netwon}$-$PN false-color image of $\eta$-Carinae field in 0.3-10.0 keV energy range from observation ID 0112560201. X- and Y- axes correspond to RA (J2000) and Dec (J2000), respectively.  \label{fig:fig10}}
\end{figure}

WR 25 was observed with \textit{XMM-Newton} 24 times from July 2000 to July 2015 with the three European Photon Imaging Camera (EPIC) instruments, \textit{viz.} MOS1, MOS2, and PN. Observations were made using various configurations of these detectors. The PN image of Eta-Carinae region  in 0.3\,-\,10.0 keV energy range is shown in Fig.~\ref{fig:fig10}. These data were reduced with SAS v15.0.0 using calibration files available in May 2017. The tasks \textit{epchain} and \textit{emchain} were used to do pipeline processing of raw EPIC Observation Data Files (ODF). List of event files was extracted using the SAS task \textit{evselect} which included selecting the good events with pattern 0 to 4 for PN and 0 to 12 for MOS data. Each observation ID was checked for pile-up using the task \textit{epatplot} but none was found to be affected by pile-up. Each dataset was checked for high background intervals and excluded from the event file wherever found. The EPIC light curves and spectra of WR 25 were extracted from on-source counts obtained from a circular region centred at the source with a radius of 30$''$. Background estimation was done from  circular regions of the same size at source-free regions surrounding the source. To apply good time intervals, dead time, exposure, point-spread function and background correction to the obtained light curves, we used the \textit{epiclccorr} task. The task \textit{especget} was used to generate the source spectrum. To calibrate the flux and energy axes, the dedicated ARF and RMF response matrices, respectively, were also calculated by this task. Backscaling of the extracted spectra was done using the task \textit{backscale}. The EPIC spectra were grouped to have minimum 20 counts per spectral bin. 

\begin{figure*}
	\centering
	\begin{subfigure}[b]{1.0\columnwidth}
		\centering
		\includegraphics[width=9.0cm,trim={0.0cm 0.5cm 0.0cm 2.5cm}]{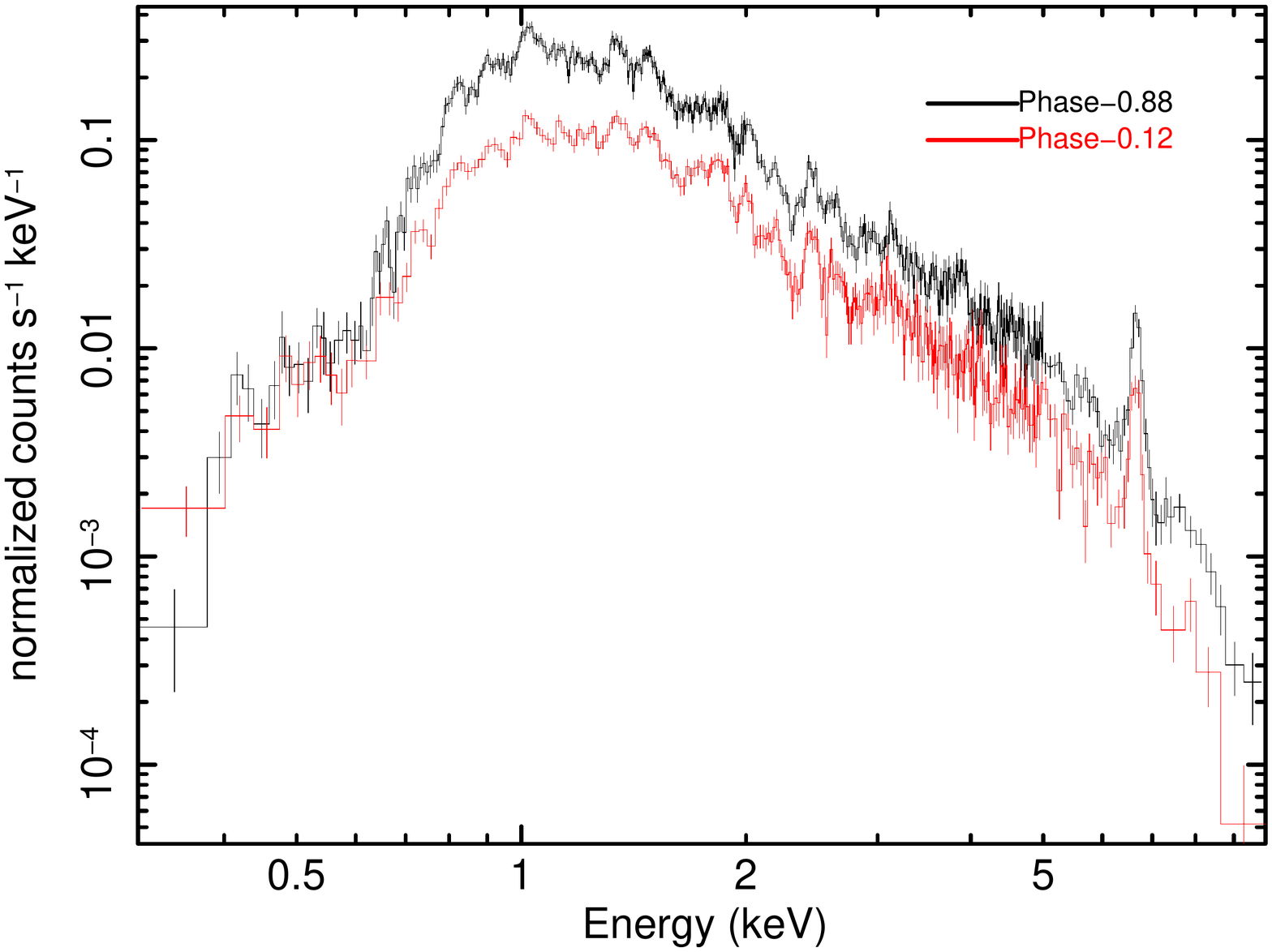}
		\caption{FI \textit{Suzaku}$-$XIS}
		\label{fig:fig5a}
	\end{subfigure}
	\hfill
	\begin{subfigure}[b]{1.0\columnwidth}
		\centering
		\includegraphics[width=9.0cm,trim={0.0cm 0.5cm 0.0cm 2.5cm}]{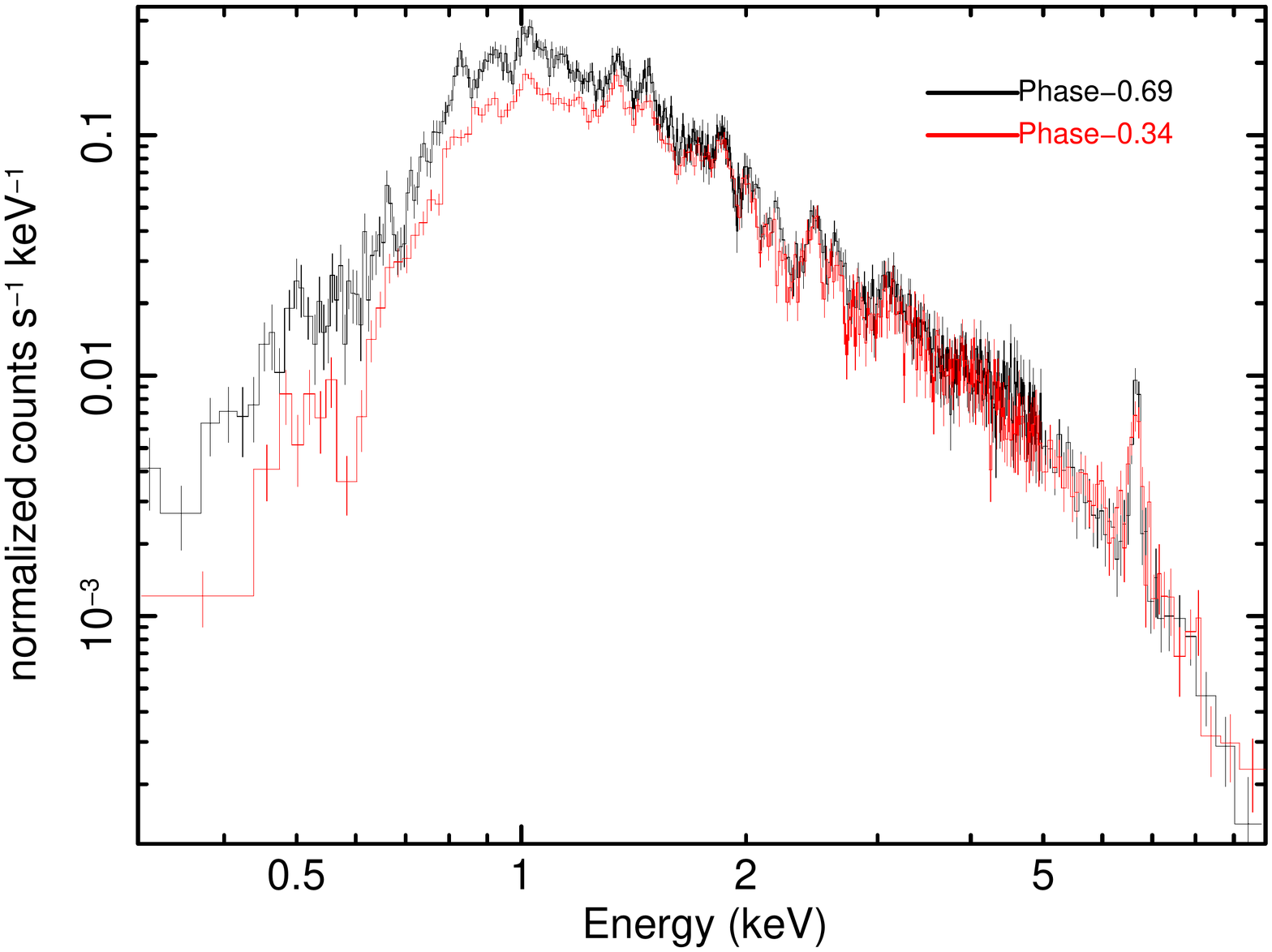}
		\caption{FI \textit{Suzaku}$-$XIS}
		\label{fig:fig5b}
	\end{subfigure}
	\hfill
	\begin{subfigure}[b]{1.0\columnwidth}
		\centering
		\includegraphics[width=9.0cm,trim={0.0cm 0.5cm 0.0cm 2.5cm}]{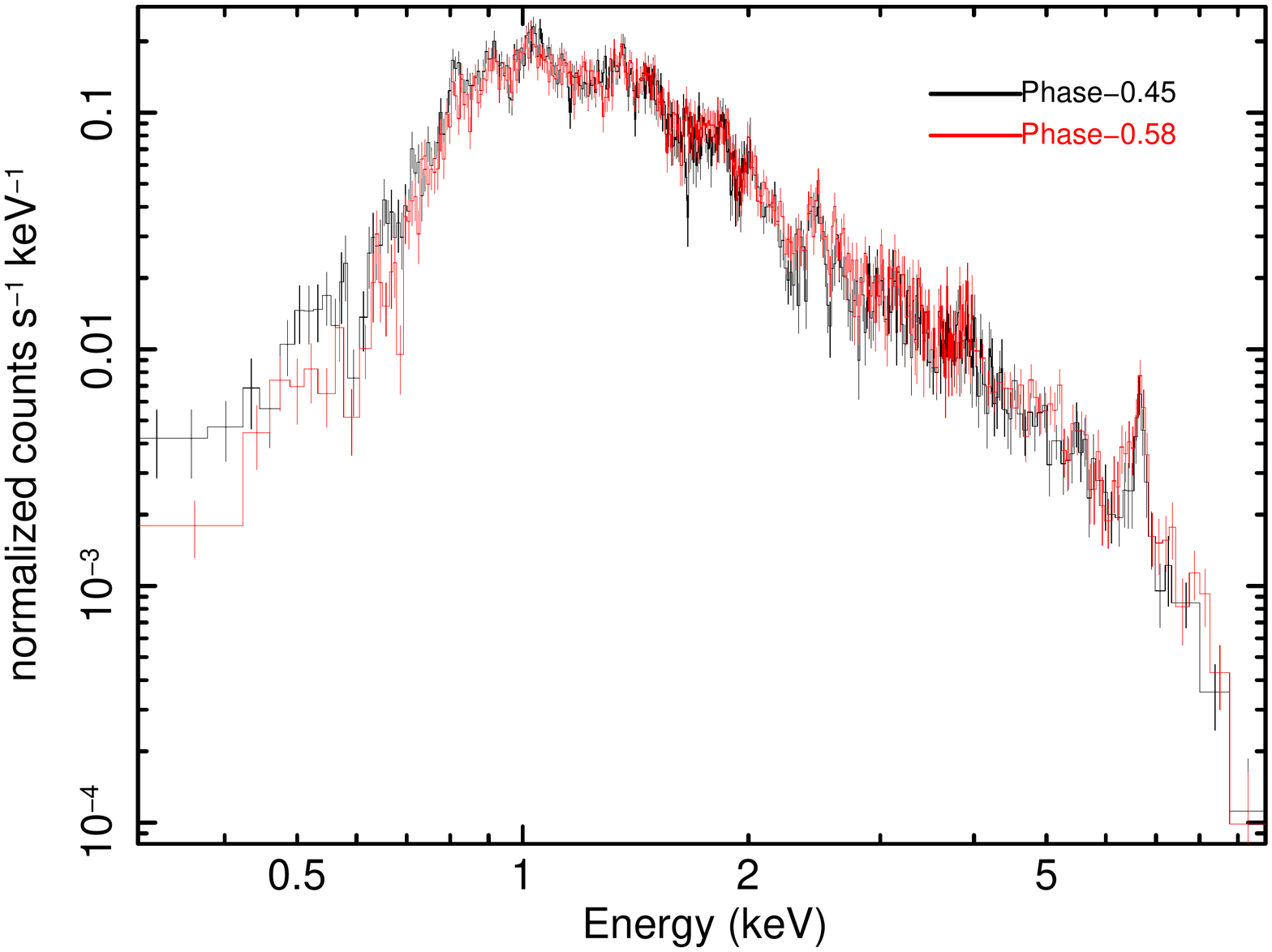}
		\caption{FI \textit{Suzaku}$-$XIS}
		\label{fig:fig5c}
	\end{subfigure}
	\hfill
	\begin{subfigure}[b]{1.0\columnwidth}
		\centering
		\includegraphics[width=9.0cm,trim={0.0cm 0.5cm 0.0cm 2.5cm}]{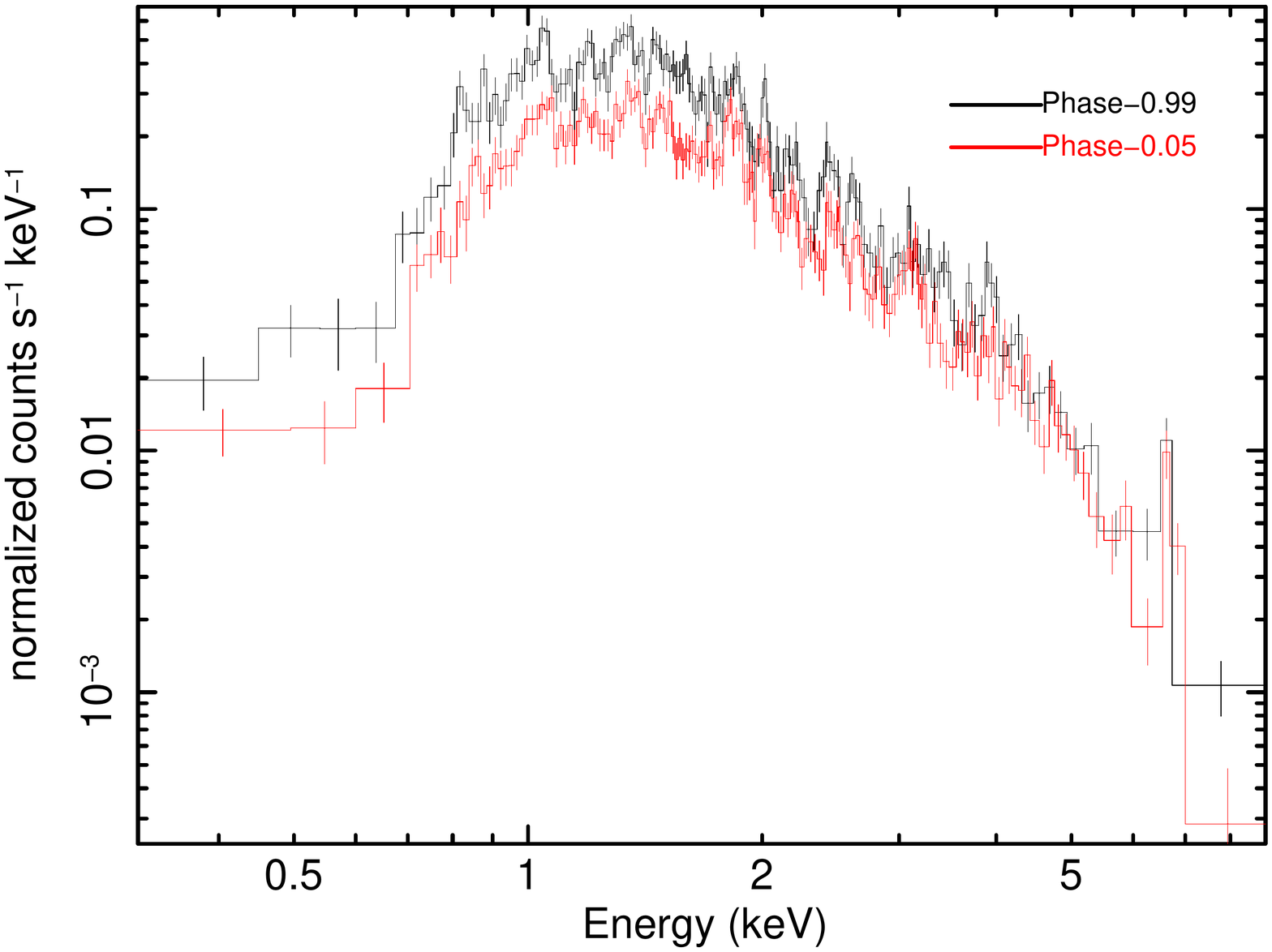}
		\caption{\textit{XMM-Newton}$-$MOS2}
		\label{fig:fig5d}
	\end{subfigure}
	\caption{FI \textit{Suzaku}$-$XIS and \textit{XMM-Newton}$-$MOS2 spectra of WR 25 at different orbital phases. In each panel the stars are at (almost) identical binary separation but either receding (phase$<$0.5) or approaching each other (phase$>$0.5).}
	\label{fig:fig5}
\end{figure*}

\begin{figure}
	\includegraphics[width=\columnwidth,trim={0.0cm 0.5cm 0.0cm 3.5cm}]{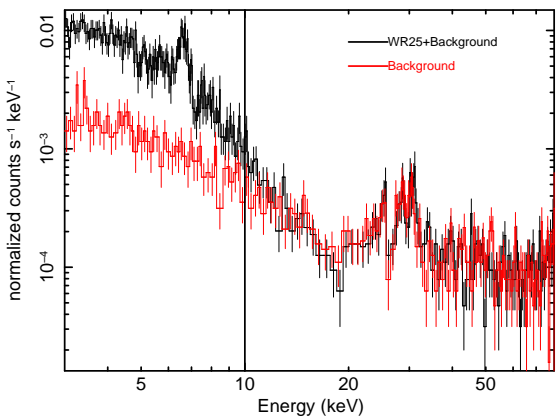}
	\caption{\textit{NuSTAR}$-$FPMA spectra of WR 25 without background correction (black line) and only background (red line) of observation ID 30002010005. \label{fig:fig4}}
\end{figure}

\subsection{\textit{NuSTAR}}
\textit{NuSTAR} observed WR 25 on 10 occasions from March 2014 to June 2016 using both the focal plane modules (FPMA and FPMB). Reduction of \textit{NuSTAR} data was done using data analysis software NuSTARDAS v1.5.1 distributed by HEASARC within HEASoft 6.18. The calibrated, cleaned, and screened event files were generated by running \textit{nupipeline} (version 0.4.4) task using \textit{NuSTAR} CALDB version 20160824 released on 2016-08-30. To extract the source and background counts, circular regions of 30$''$ radius at a source and nearby surrounding source-free regions, respectively, on the same detector were selected. The size of the chosen source region is of the order of the HPD ($\sim$1$\arcmin$). Light curve and spectrum in 3$-$79 keV energy range and corresponding response files were extracted using \textit{nuproducts} package within NuSTARDAS with time binning of 10 sec. In order to correct the arrival times of the X-ray photons, the barycentric correction was applied to the background subtracted light curves using the task \textit{barycorr}. Spectra were grouped to have 10 counts per spectral bin using the task GRPPHA. Spectra extracted from FPMA and FPMB of individual dataset were fitted simultaneously. 

\begin{figure}
	\includegraphics[width=\columnwidth,scale=0.1,trim={1.9cm 1.8cm 7cm 18.2cm}]
	{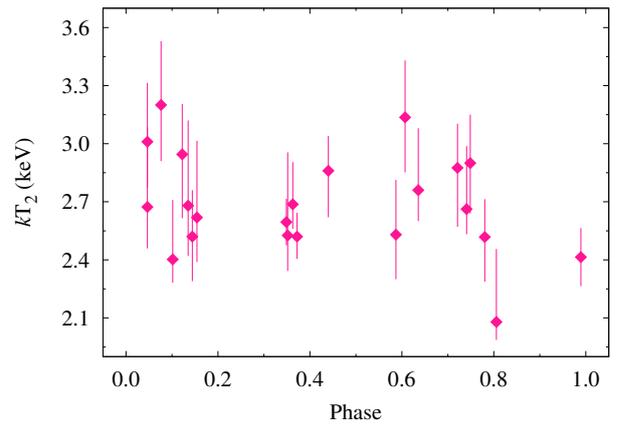}
	\caption{Variation of plasma temperature corresponding to hot component ($kT_{2}$) for \textit{XMM-Newton}$-$EPIC data.\label{fig:fig11}}
\end{figure}

\begin{figure*}
    \centering
    \begin{subfigure}[b]{0.33\textwidth}
        \centering
        \includegraphics[width=7.0cm,scale=1.0,trim={1cm 2.0cm 5.5cm 4.5cm}]{{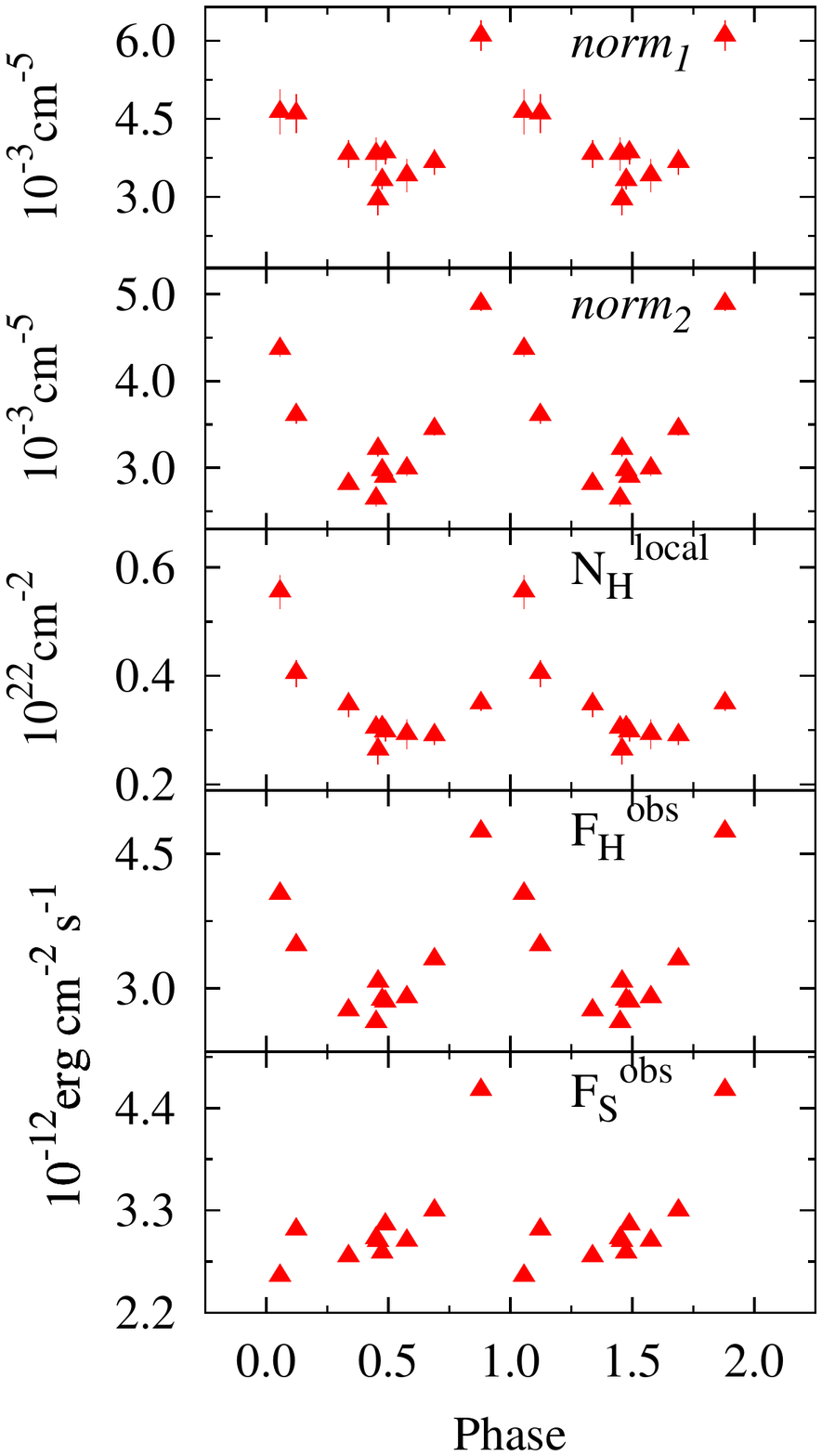}}
        \caption{\textit{Suzaku}$-$XIS}
        \label{fig:fig7a}
    \end{subfigure}
    \hfill
    \begin{subfigure}[b]{0.33\textwidth}
        \centering
        \includegraphics[width=7.0cm,scale=1,trim={1.0cm 2.0cm 5.5cm 4.5cm}]{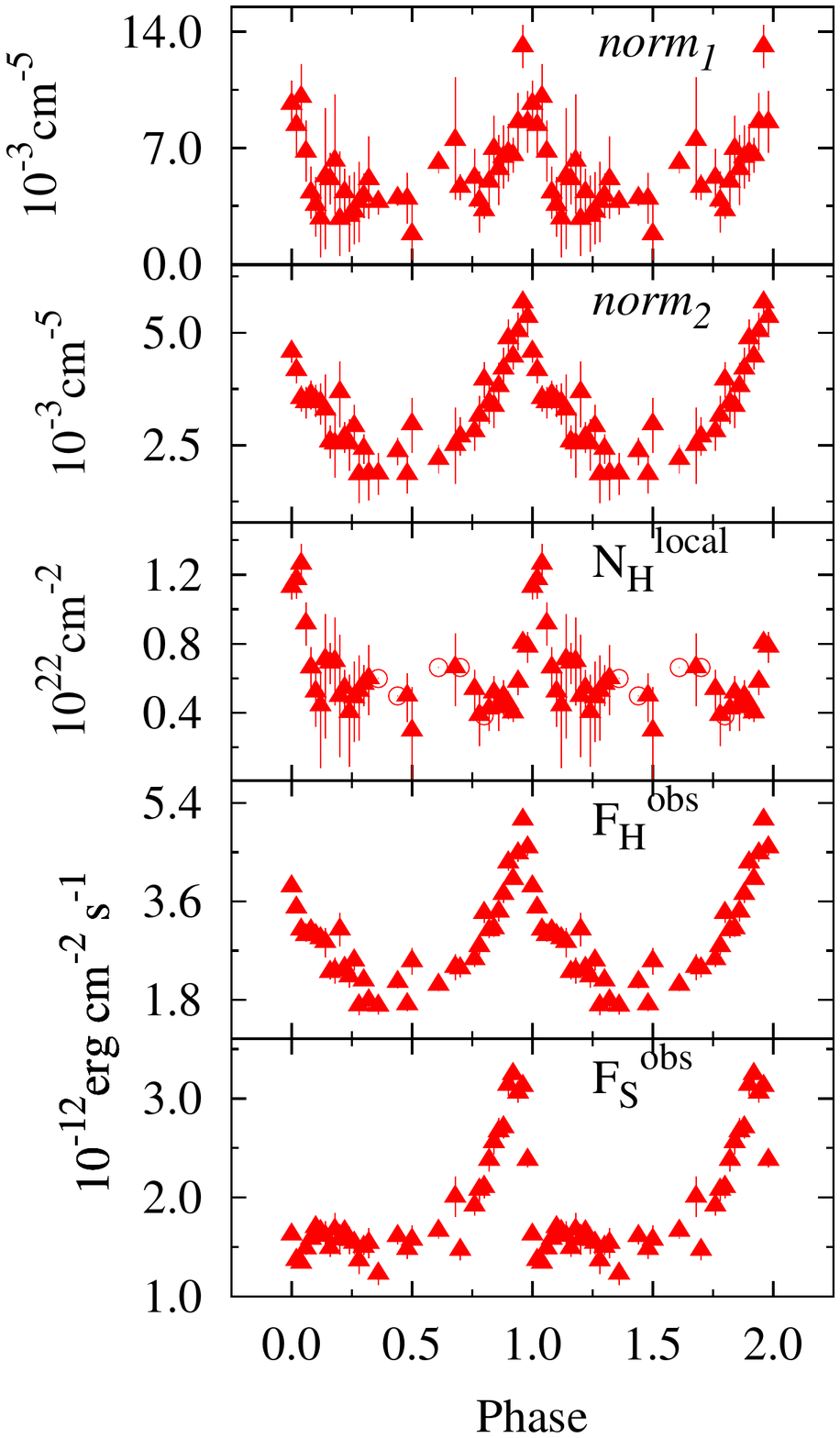}
        \caption{\textit{Swift}$-$XRT}
        \label{fig:fig7b}
    \end{subfigure}
    \hfill
    \begin{subfigure}[b]{0.33\textwidth}
        \centering
        \includegraphics[width=7.0cm,scale=1,trim={1cm 2.0cm 5.5cm 4.5cm}]{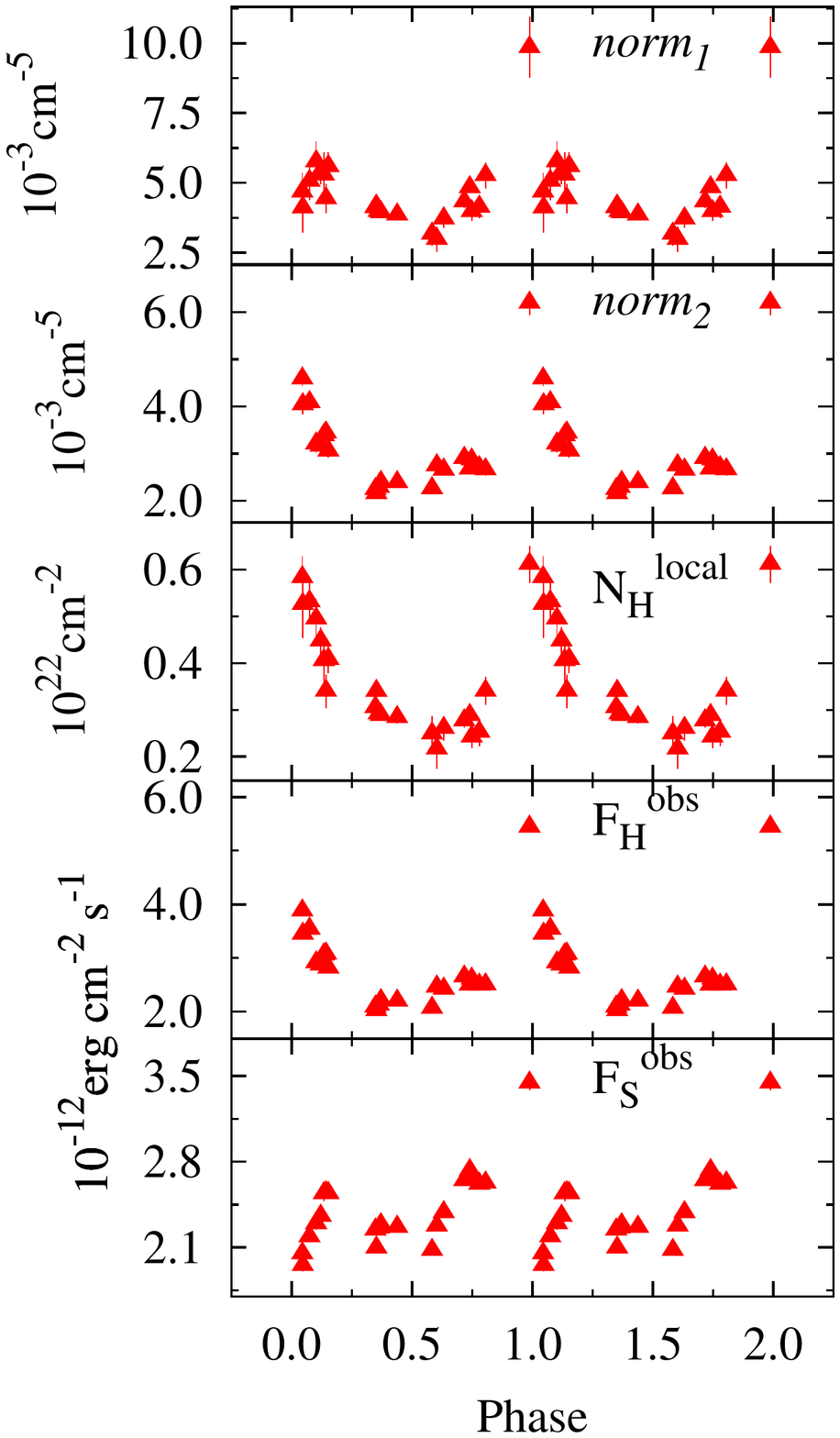}
        \caption{\textit{XMM-Newton}$-$EPIC}
        \label{fig:fig7c}
    \end{subfigure}
    \caption{Spectral parameters as a function of orbital phase as observed by (a) \textit{Suzaku}$-$XIS,  (b) \textit{Swift}$-$XRT, and (c) \textit{XMM-Newton}$-$EPIC. Open circles in the middle panel of the middle figure corresponds to the fixed values of $N_H^{local}$ to those of the nearby phase bins. 
}
    \label{fig:fig7}
\end{figure*}

\begin{figure}
\includegraphics[width=\columnwidth,trim={0.0cm 2.8cm 0.0cm 2.6cm}]{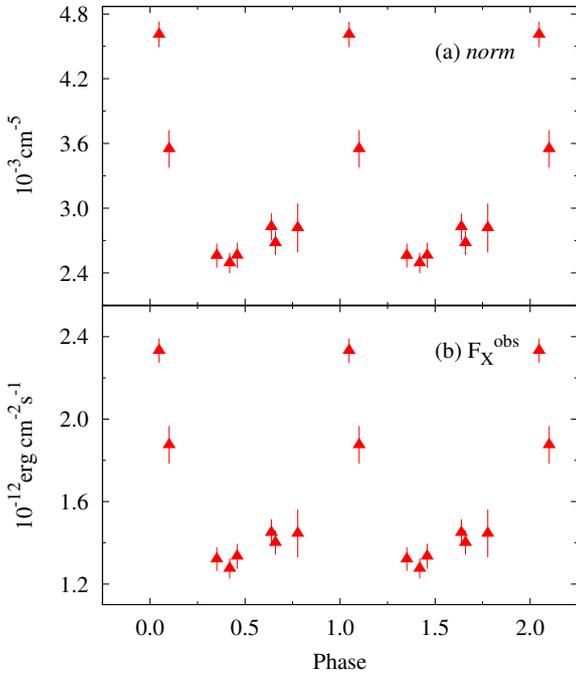}
\caption{Normalization constant (\textit{norm}) and observed X-ray flux ($F_{X}^{obs}$) as a function of orbital phase for \textit{NuSTAR}$-$FPMs data.\label{fig:fig6}}
\end{figure}

\section{The Spectra}\label{spectra}
\subsection{Look at the X-ray spectra of WR 25}\label{lookatspectra}
The X-ray spectrum below 10.0\,keV presents the typical features of an optically thin thermal plasma emission at high temperature (above 10$^6$\,K), as expected for a colliding wind massive binary \citep[see e.g.][]{Pandey2014}.
 
The FI \textit{Suzaku}$-$XIS spectrum of WR 25 at different orbital phases (but at almost same binary separation) are shown in Figs.~\,\ref{fig:fig5a}, \ref{fig:fig5b}, and \ref{fig:fig5c}. The two spectra in each of the figure show the difference in the source flux when the two components of the binary system move towards and away from each other. Fig.~\ref{fig:fig5a} shows the source spectra before ($\phi$ = 0.88, black in colour) and after ($\phi$ = 0.12, red in colour) periastron passage. The system is brighter in X-rays just before periastron passage than after it at all the energies. Similarly the \textit{XMM-Newton}$-$MOS2 spectra of WR 25, as shown in Fig.~\ref{fig:fig5d}, display the same behavior. However, as the two components move towards apastron this difference keeps on decreasing (Fig.~\ref{fig:fig5b}) and vanishes close to apastron (Fig.~\ref{fig:fig5c}).     

The \textit{NuSTAR}$-$FPMA spectrum of WR 25 without background subtraction (black in colour) as well as the background spectrum (red in colour) in 3.0$-$79.0 keV energy range are shown in Fig.~\ref{fig:fig4}. It is evident that after 10.0 keV, both spectra are almost identical which confirms that the source counts are significant only in 3.0$-$10.0 keV energy range. Background photons are dominating in the extracted spectrum beyond 10.0 keV. The lack of hard X-ray emission is discussed in Sect.\,\ref{nonthermal}. Therefore, the \textit{NuSTAR} spectrum of WR 25 was considered in 3.0$-$10.0 keV energy range only for further spectral analysis.

\subsection{X-ray spectral analysis}
The fitting of X-ray spectra in 0.3$-$10.0 keV energy range was done using the models of the Astrophysical Plasma Emission Code (APEC, \citealt{Smith2001}) in the X-ray spectral fitting package \textit{xspec} \citep{Arnaud1996} version 12.9.0i. The form of the model used was \textit{$phabs(ism)*phabs(local) (vapec+vapec)$}. The component \textit{phabs} used to model the interstellar as well as local absorption effects  uses the values of different elemental abundances according to \citet{Anders1989}. A similar model was also used by  \citet{Pandey2014} for the spectral analysis of WR 25. However, \citet{Pandey2014} used \textit{wabs} instead of \textit{phabs}. Since the X-ray spectra from \textit{XMM-Newton} have best photon statistics than the others, therefore, firstly, all the parameters excluding N$_H^{ISM}$  were free in the spectral fitting. Temperature corresponding to cool component was found to be constant around its mean value of 0.628 keV.  Phased variation of temperatures corresponding to hot component \textit{(k}T$_2$) is shown in  Figure \ref{fig:fig11}, which is also found to be constant within 1$\sigma$ level around the mean value of 2.75 keV. The model parameters derived from the current spectral fitting which are based on the latest calibration, software, and larger data sets are found to be similar to that found by  \citet{Pandey2014}. Therefore, the values of the interstellar equivalent H-column density ($N_H^{ISM}$), as well as the temperature and abundances for the two thermal components, were kept fixed at those obtained by \citet{Pandey2014}. Only the local equivalent H-column density ($N_H^{local}$), as well as the normalization constants for the two temperature components, were kept free. $\chi^{2}$ minimization method was used to best fit the model to the data. The observed (not corrected for any absorption effects, $F^{obs}$), ISM corrected ($F^{ism}$),  and intrinsic (corrected for both ISM and local wind absorption, $F^{int}$) X-ray fluxes of WR 25 in soft (0.3$-$2.0 keV, $F_{S}$), hard (2.0$-$10.0 keV, $F_{H}$), and 3.0$-$10.0 keV ($F_{X}$) energy bands were derived using the model \textit{cflux} in \textit{xspec}. The normalization constants, $norm_{1}$ and $norm_{2}$, corresponding to the cool and hot plasma components, respectively, were also estimated. The method used for fitting of the spectra from each satellite, with this model, is explained as follows:

\subsubsection{Suzaku} The BI and FI \textit{Suzaku}$-$XIS spectra of WR 25 were fitted simultaneously in 0.7$-$10.0 keV energy range due to poor SNR of data below 0.7\,keV. The spectra obtained from each dataset were fitted individually. The spectral parameters obtained after the spectral fitting (\textit{viz.} $norm_{1}$, $norm_{2}$, $N_{H}^{local}$, $F_{S}^{obs}$, $F_{H}^{obs}$, $F^{ism}_{S}$, $F^{ism}_{H}$, $F_{S}^{int}$, and $F_{H}^{int}$) are given in Table~\ref{tab:spectra} and are plotted as a function of orbital phase in Fig.~\ref{fig:fig7a}. 

\subsubsection{Swift} The fitting of the \textit{Swift}$-$XRT spectra was done in the 0.5$-$10.0 keV energy range. Since the signal-to-noise ratio of individual \textit{Swift} spectra is very low, we decided to simultaneously fit the spectra lying within a phase interval of $\leqslant$ 0.02. But still, we were not able to fit certain phase bins spectra due to very poor SNR. To solve this problem, we had to simultaneously fit those spectra with those of nearby phase bins. In addition, some orbital phase bins were not covered by observations. As a result, we defined phase bins 1 to 36 (introduced in the Table~ \ref{tab:spectra}) and a simultaneous fitting of the spectra taken in the same phase bin was performed. The values of the different spectral parameters obtained are given in Table~\ref{tab:spectra} and their variation with orbital phase is shown in Fig.~\ref{fig:fig7b}. For the five phase bins, \textit{viz.} 32, 33, 34, 35, and 36, we had to fix the $N_H^{local}$ values to those obtained by the spectral fitting of the nearby phase bins. The open circles in the middle panel of Fig.~\ref{fig:fig7b} mark these values.

\subsubsection{XMM-Newton} The spectra obtained from MOS1, MOS2, and PN detectors of the same dataset were fitted simultaneously in the 0.3$-$10.0 keV energy range. The spectra obtained from observation IDs 0145740201 and 0145740301 were added since these were observed on the same day. Similarly, spectra from IDs 0145740401 and 0145740501 were also added. Other spectra from different datasets were fitted individually.   The spectral parameters obtained after fitting of EPIC spectra are given in Table~\ref{tab:spectra}. These parameters are plotted as a function of orbital phase in Fig.~\ref{fig:fig7c}.

\subsubsection{NuSTAR} The \textit{NuSTAR} spectrum was fitted in the 3.0$-$10.0 keV energy range only for the reasons mentioned in the above sub-section. Since the \textit{NuSTAR} spectrum does not include the soft energy range, it was fitted using the single temperature component with $kT=2.75$ keV. The form of the model used was \textit{$phabs(ism)*phabs(local)*vapec$}. The spectrum obtained by the detectors FPMA and FPMB of the same observation dataset were fitted simultaneously. Additionally, the spectrum observed within a phase difference of $\leqslant$ 0.02 were also fitted jointly. Since the soft energy range is missing in \textit{NuSTAR} spectra, $N_H^{local}$ could not be determined through the standard fitting procedure. We had to freeze the $N_H^{local}$ values, for \textit{NuSTAR} spectra, to those obtained by the fitting of the very close-by phases spectra from other satellites data. The normalization constant ($norm$), and the resulting observed ($F_{X}^{obs}$), ISM corrected ($F_{X}^{ism}$) as well as intrinsic ($F_{X}^{int}$) X-ray fluxes were determined and are given in Table~\ref{tab:nustar}. The variation of $norm$ and $F_{X}^{obs}$ with orbital phase is shown in Fig.~\ref{fig:fig6}.


\begin{table*} 
	\caption{Best fit parameters obtained from \textit{Suzaku}, \textit{Swift}, and \textit{XMM-Newton} spectral fitting of WR 25 }
	\setlength{\tabcolsep}{4pt}
	\begin{tabular}{c c c c c c c c c c c c}
		\toprule[1.5pt] 
		Obs. ID / & $\phi$  & $N_{H}^{local}$     & $norm_{1}$        & $norm_{2}$          & $F^{obs}_{S}$ & $F^{obs}_{H}$ &  $F^{ism}_{S}$ & $F^{ism}_{H}$ & $F^{int}_{S}$ & $F^{int}_{H}$  & $\chi^{2}_{\nu} (dof)$ \\ 
	Phase bin &         & ($10^{22}$ cm$^{-2}$) & \multicolumn{2}{c}{($10^{-3}$ cm$^{-5}$)}  & \multicolumn{6}{c}{($10^{-12}$ erg cm$^{-2}$ s$^{-1}$)} & \\ 
		\midrule \midrule 	
		\multicolumn{12}{c}{\textit{Suzaku}} \\
		100012010&0.689&$0.29^{+0.02}_{-0.02}$&$3.67^{+0.25}_{-0.25}$&$3.45^{+0.07}_{-0.07}$&$3.29^{+0.03}_{-0.03}$&$3.32^{+0.03}_{-0.03}$&$4.68^{+0.04}_{-0.04}$ &$3.39^{+0.03}_{-0.03}$ &$12.23^{+0.11}_{-0.11}$&$3.55^{+0.03}_{-0.03}$&1.32 (775)\\
		100045010&0.450&$0.30^{+0.02}_{-0.02}$&$3.82^{+0.32}_{-0.32}$&$2.64^{+0.09}_{-0.09}$&$2.99^{+0.04}_{-0.04}$&$2.62^{+0.03}_{-0.04}$&$4.27^{+0.06}_{-0.06}$ &$2.69^{+0.04}_{-0.04}$ &$11.72^{+0.16}_{-0.16}$&$2.81^{+0.04}_{-0.04}$&1.39 (753)\\
		402039010&0.879&$0.35^{+0.01}_{-0.01}$&$6.09^{+0.29}_{-0.29}$&$4.89^{+0.08}_{-0.08}$&$4.59^{+0.03}_{-0.03}$&$4.74^{+0.03}_{-0.03}$&$6.39^{+0.05}_{-0.05}$ &$4.85^{+0.03}_{-0.03}$ &$19.42^{+0.14}_{-0.14}$&$5.12^{+0.04}_{-0.04}$&1.46 (730)\\
		403035010&0.577&$0.29^{+0.03}_{-0.03}$&$3.41^{+0.32}_{-0.32}$&$2.99^{+0.08}_{-0.08}$&$2.97^{+0.04}_{-0.04}$&$2.89^{+0.04}_{-0.04}$&$4.22^{+0.05}_{-0.05}$ &$2.97^{+0.04}_{-0.04}$ &$11.13^{+0.14}_{-0.14}$&$3.09^{+0.04}_{-0.04}$&1.11 (634)\\
		
		..& .. & .. & .. & ..& .. & .. &.. & .. & ..& .. &.. \\
		..& .. & .. & .. & ..& .. & .. &.. & .. & ..& .. &.. \\
		\midrule
		\label{tab:spectra}	
	\end{tabular}
	\begin{tablenotes}
		\small
		\item \textbf{Notes:} The fitted model has the form \textit{$phabs(ism)*phabs(local)(vapec+vapec)$}, with $N_{H}^{ISM}$ fixed to $3.7\times10^{21}$ cm$^{-2}$ and the temperatures fixed to 0.628 and 2.75 keV. Abundances are given in Table 4 of \citet{Pandey2014}.  $\chi_{\nu}^{2}$ is the reduced $\chi^2$ and \textit{dof} is degrees of freedom. Errors quoted on different parameters refer to 90 per cent confidence level. The $N_{H}^{local}$ values for 32 to 36 phase bins of \textit{Swift} were fixed to those of the nearby phase bins. The \textit{XMM-Newton} spectra obtained from observation IDs 0145740201 and 0145740301 as well as 0145740401 add 0145740501 were added. Full table is available online.
	\end{tablenotes}
\end{table*}

\begin{table*}
	\centering
	\caption{Best fit parameters obtained from \textit{NuSTAR} spectral fitting of WR 25 }
	\begin{tabular}{c c c c c c c c}
		\toprule[1.5pt]
		Obs. & $\phi$  & $N_{H}^{local}$     & $norm$       & $F^{obs}_{X}$ & $F^{ism}_{X}$ & $F^{int}_{X}$  & $\chi^{2}_{\nu} (dof)$ \\ 
		ID  &                             & ($10^{22}$ cm$^{-2}$) & ($10^{-3}$ cm$^{-5}$)  & \multicolumn{3}{c}{($10^{-12}$ erg cm$^{-2}$ s$^{-1}$)} &   \\ 
		\midrule \midrule 	
		30002010002 & 0.777    &0.29 &$ 2.82^{+0.23}_{-0.23}$ & $1.45^{+0.12}_{-0.12}$ &$1.48^{+0.12}_{-0.12}$  & $1.50^{+0.12}_{-0.12}$ &  1.26 (31) \\
		\midrule
		30002010005 & \multirow{2}{*}{0.048}   &\multirow{2}{*}{0.99} & \multirow{2}{*}{$4.61^{+0.12}_{-0.12}$} &  \multirow{2}{*}{$2.33^{+0.06}_{-0.06}$} &\multirow{2}{*}{$2.38^{+0.06}_{-0.06}$}  &\multirow{2}{*}{$2.52^{+0.06}_{-0.06}$} &\multirow{2}{*}{1.34 (63)} \\
		30101005002 &	  &            &     &		        &              &         &  \\
		\midrule
		30002040004 & 0.351    &0.34 &$ 2.56^{+0.11}_{-0.11}$ & $1.32^{+0.06}_{-0.06}$  &$1.35^{+0.06}_{-0.06}$  & $1.38^{+0.06}_{-0.06}$ &  1.04 (31) \\
		\midrule
		30002010007 & \multirow{2}{*}{0.419}      &\multirow{2}{*}{0.32} & \multirow{2}{*}{$ 2.49^{+0.09}_{-0.09}$} & \multirow{2}{*}{$1.28^{+0.05}_{-0.05}$}  &\multirow{2}{*}{$1.30^{+0.05}_{-0.05}$} &\multirow{2}{*}{$1.33^{+0.05}_{-0.05}$} &\multirow{2}{*}{1.35 (63)} \\
		30002010008 &	  &            &     &		        &               &         & \\
		\midrule
		30002010010 & 0.458    &0.26 &$ 2.56^{+0.12}_{-0.12}$  &$1.33^{+0.06}_{-0.06}$ &$1.36^{+0.06}_{-0.06}$  & $1.39^{+0.06}_{-0.06}$ &  1.03 (31) \\ 
		30002010012 & 0.639    &0.28 &$ 2.83^{+0.12}_{-0.12}$  &$1.45^{+0.06}_{-0.06}$ &$1.48^{+0.07}_{-0.07}$  & $1.51^{+0.07}_{-0.07}$ &  1.69 (31) \\ 
		30002040002 & 0.101    &0.45 &$ 3.55^{+0.17}_{-0.17}$  &$1.87^{+0.09}_{-0.09}$ &$1.92^{+0.09}_{-0.09}$  & $1.97^{+0.09}_{-0.09}$ &  0.82 (31) \\ 
		30201030002 & 0.660    &0.32 &$ 2.68^{+0.11}_{-0.11}$  &$1.40^{+0.06}_{-0.06}$ &$1.43^{+0.06}_{-0.06}$  & $1.46^{+0.06}_{-0.06}$ &  1.49 (31) \\    [1ex]	
		\bottomrule[1.5pt]
		\label{tab:nustar}	
	\end{tabular}
	\begin{tablenotes}
		\small
		\item \textbf{Notes:} The fitted model has the form \textit{$phabs(ism)*phabs(local)*vapec$}, with $N_{H}^{ISM}$ fixed to $3.7\times10^{21}$ cm$^{-2}$ and the temperature fixed to 2.75 keV. Abundances are given in Table 4 of \citet{Pandey2014}. $\chi_{\nu}^{2}$ is the reduced $\chi^2$  and  \textit{dof} is degrees of freedom.  Errors quoted on different parameters refer to 90 per cent confidence level. The spectra obtained from observation IDs 30002010005 and 30101005002 as well as 30002010007 and 30002010008 were fitted simultaneously. The $N_{H}^{local}$ values for \textit{NuSTAR} spectra were fixed to those obtained by the spectral fitting of the very close-by phases spectra from other observatories data. 
	\end{tablenotes}
\end{table*}

\section{X-ray light curves analysis}\label{lc}
\begin{figure}
\includegraphics[width=\columnwidth,trim={0.0cm 2.0cm 0.0cm 0.5cm}]{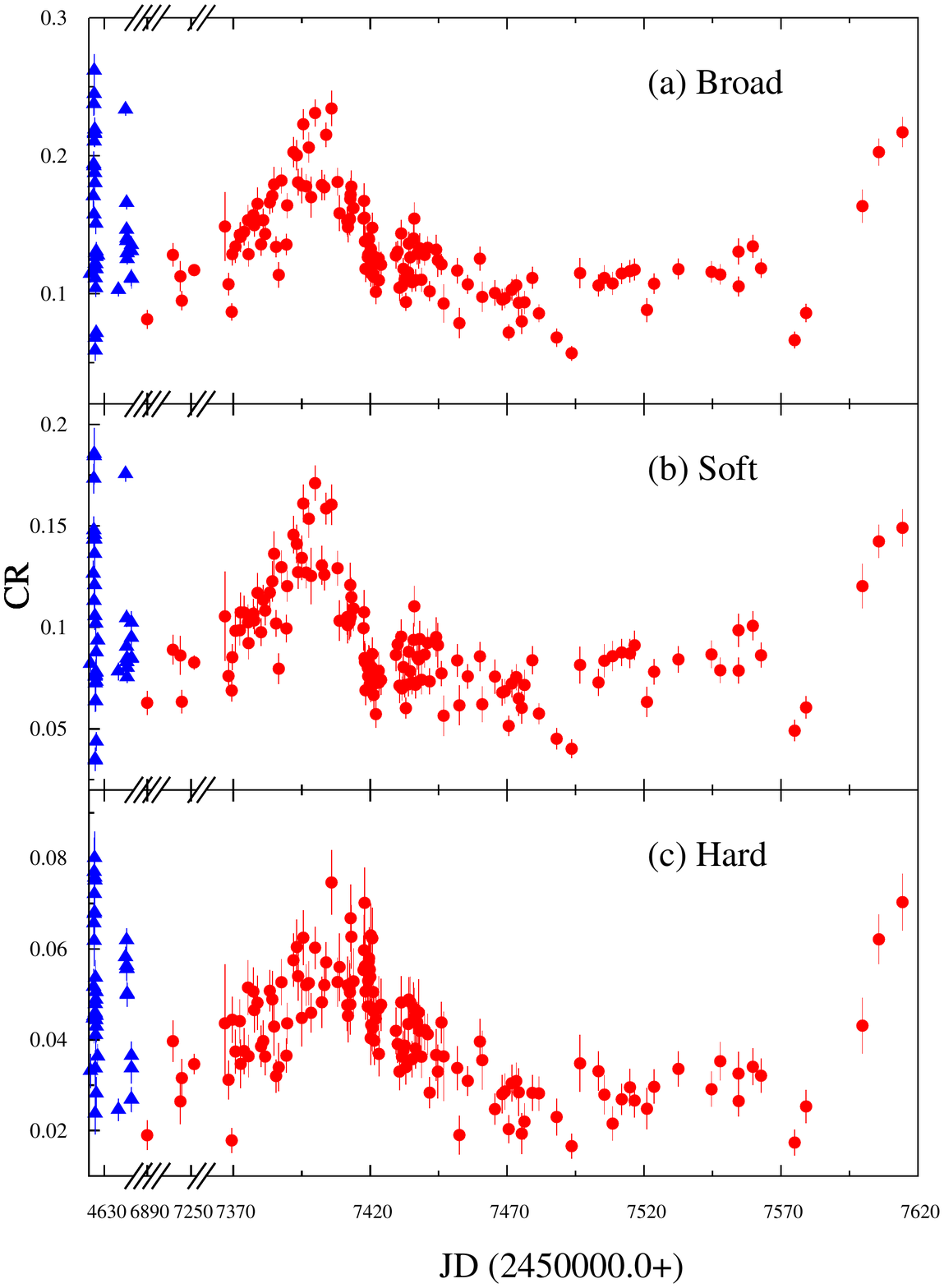}
\caption{X-ray light curves in broad (0.3$-$10.0 keV), soft (0.3$-$2.0 keV), and hard (2.0$-$10.0 keV) energy bands as observed by \textit{Swift}$-$XRT. Blue triangles mark the data observed during December 2007 to June 2009 and the red filled circles correspond to August 2014 to August 2016 observations. 
\label{fig:fig1}}
\end{figure}

The background subtracted X-ray light curves as observed by \textit{Swift}-XRT in broad (0.3$-$10.0 keV), soft (0.3$-$2.0 keV), and hard (2.0$-$10.0 keV) energy bands are shown in Fig.~\ref{fig:fig1}, where each data point represents an average count rate of the corresponding observation. Blue triangles mark the data observed during December 2007 to June 2009 which was studied by \citet{Pandey2014}. However, red filled circles corresponds to the data observed continuously from August 2014 to August 2016. A time span of 250 days is covered between JD 2457370.0 and 2457620.0. The time period of this binary system is estimated to be 207.85$\pm$0.02 days by \citet{Gamen2006} from radial velocity measurements of WR 25 using N IV $\lambda$4058 emission line. \citet{Pandey2014} also found a period of 207.5\,$\pm$\,3.4 days on the basis of X-ray light curves. The present 250 days continuous monitoring of WR 25 covers more than one orbital cycle in Fig.~\ref{fig:fig1}. Therefore, count rate initially increased (at around JD = 2457370.0) to the maximum and then decreased (at around JD = 2457495.0) to the minimum and then again increased towards the end of the light curve in all the energy bands mentioned here. However, the change in the hard energy band is not as steep as that in the soft energy band. 

The long-term observation of WR 25 also enabled us to determine its orbital period more accurately than the previous studies. Lomb-Scargle periodogram \citep{Lomb1976,Scargle1982} was used to perform the period analysis from \textit{Swift} light curves. The peak with the highest power lies at frequency 0.00481\,$\pm$\,0.00005 cycles day$^{-1}$ in broad energy band. This corresponds to 207.8\,$\pm$\,3.4, days orbital period which is consistent with the previous findings also. The power spectra of \textit{XMM-Newton} light curves also showed a similar orbital period of $208.3\pm2.2$ days.

The light curves from other observatories (\textit{NuSTAR}, \textit{Suzaku}, and \textit{XMM-Newton}) data are not shown here because none of them presents a continuous monitoring of the source. Only a few observations of WR\,25 were made at random orbital phases and hence no regular pattern was visible in their light curves.

\subsection{Folded X-ray light curves}
\begin{figure*}
    \centering
    \begin{subfigure}[b]{0.33\textwidth}
        \centering
        \includegraphics[width=6.4cm]{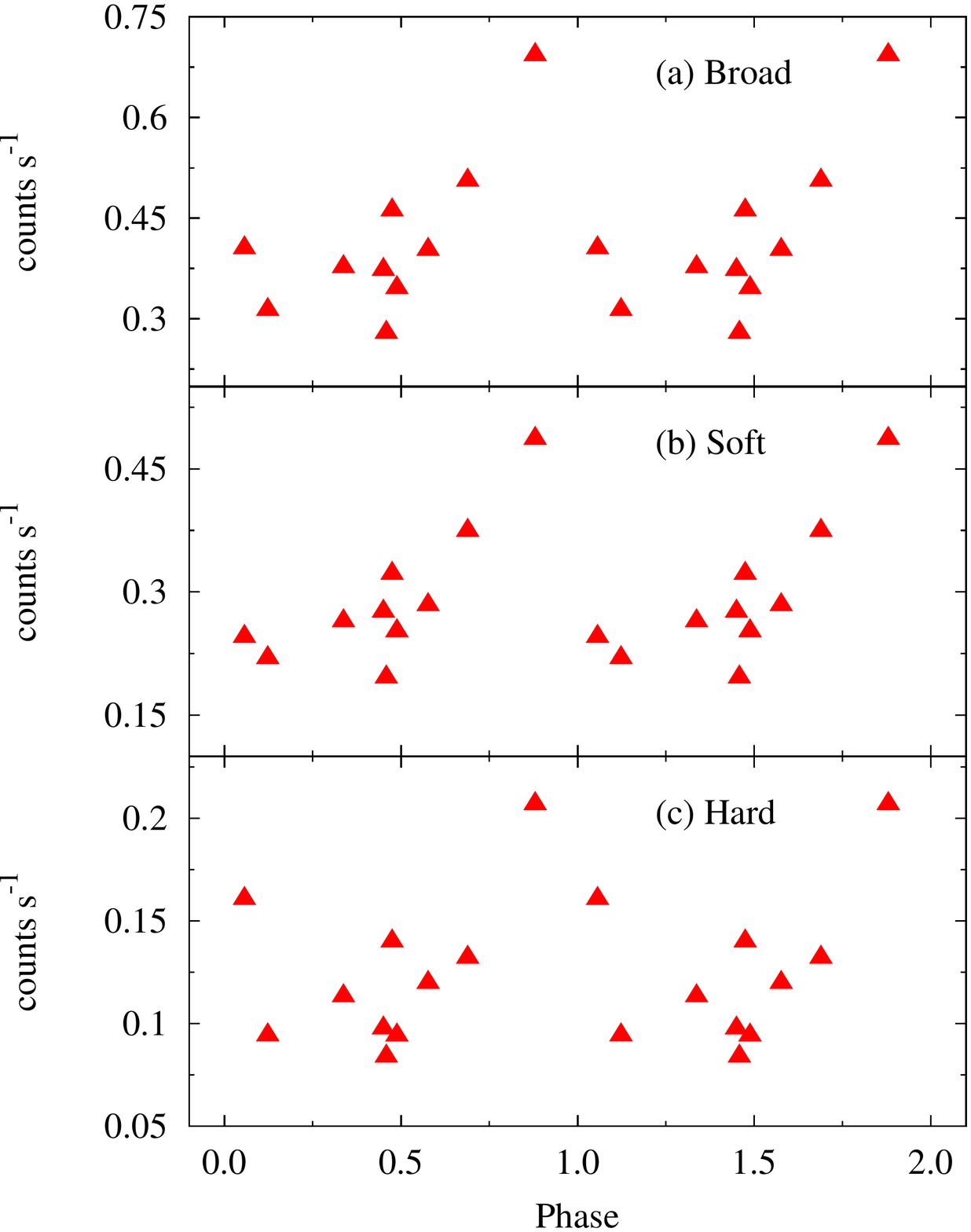}
        \caption{FI \textit{Suzaku}$-$XIS}
        \label{fig:fig3a}
    \end{subfigure}
   \hfill
    \begin{subfigure}[b]{0.35\textwidth}
        \centering
        \includegraphics[width=6.4cm]
        {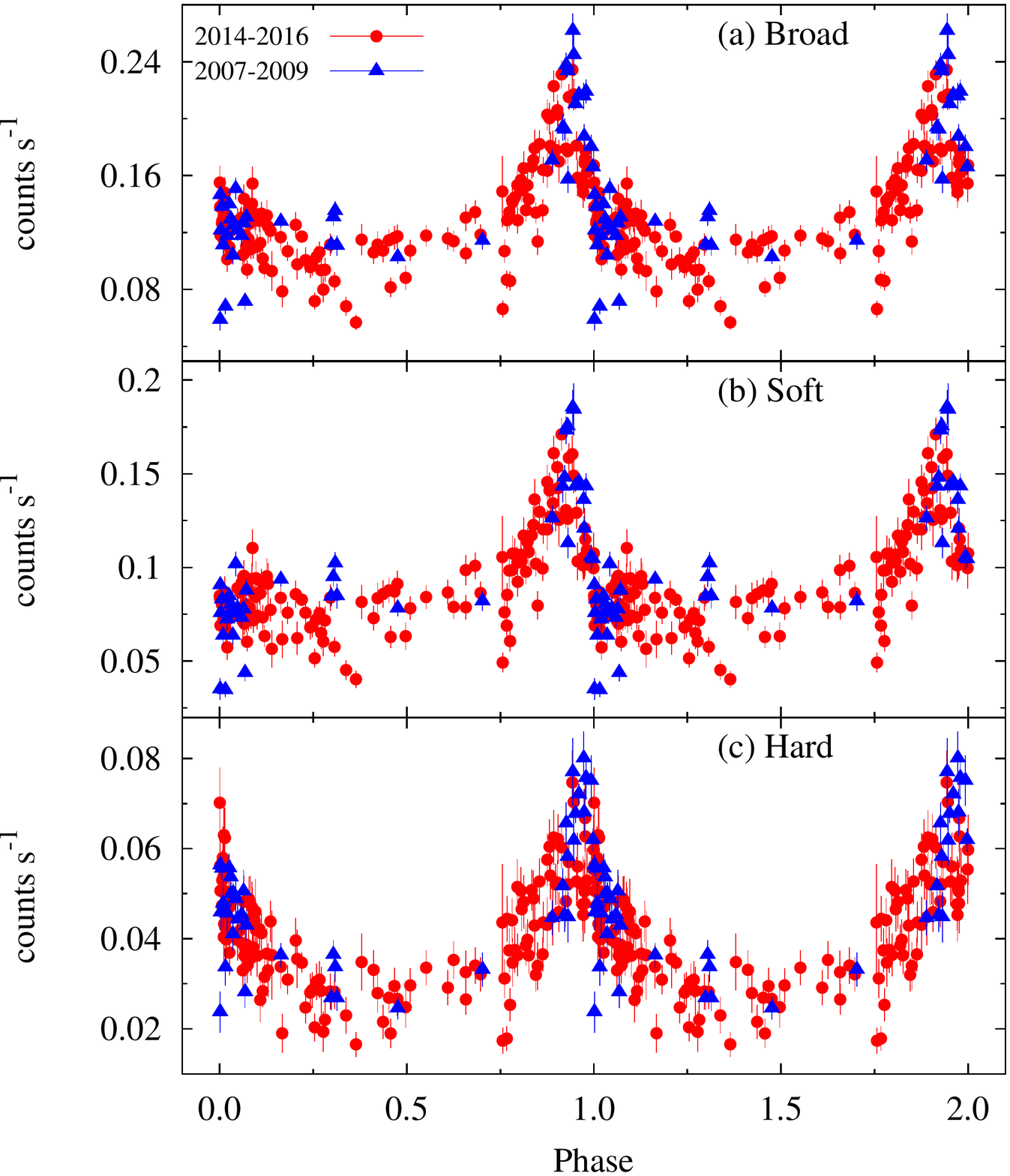}
        \caption{\textit{Swift}$-$XRT}
        \label{fig:fig3b}
    \end{subfigure}
   \hfill
    \begin{subfigure}[b]{0.31\textwidth}
        \centering
        \includegraphics[width=6.4cm]{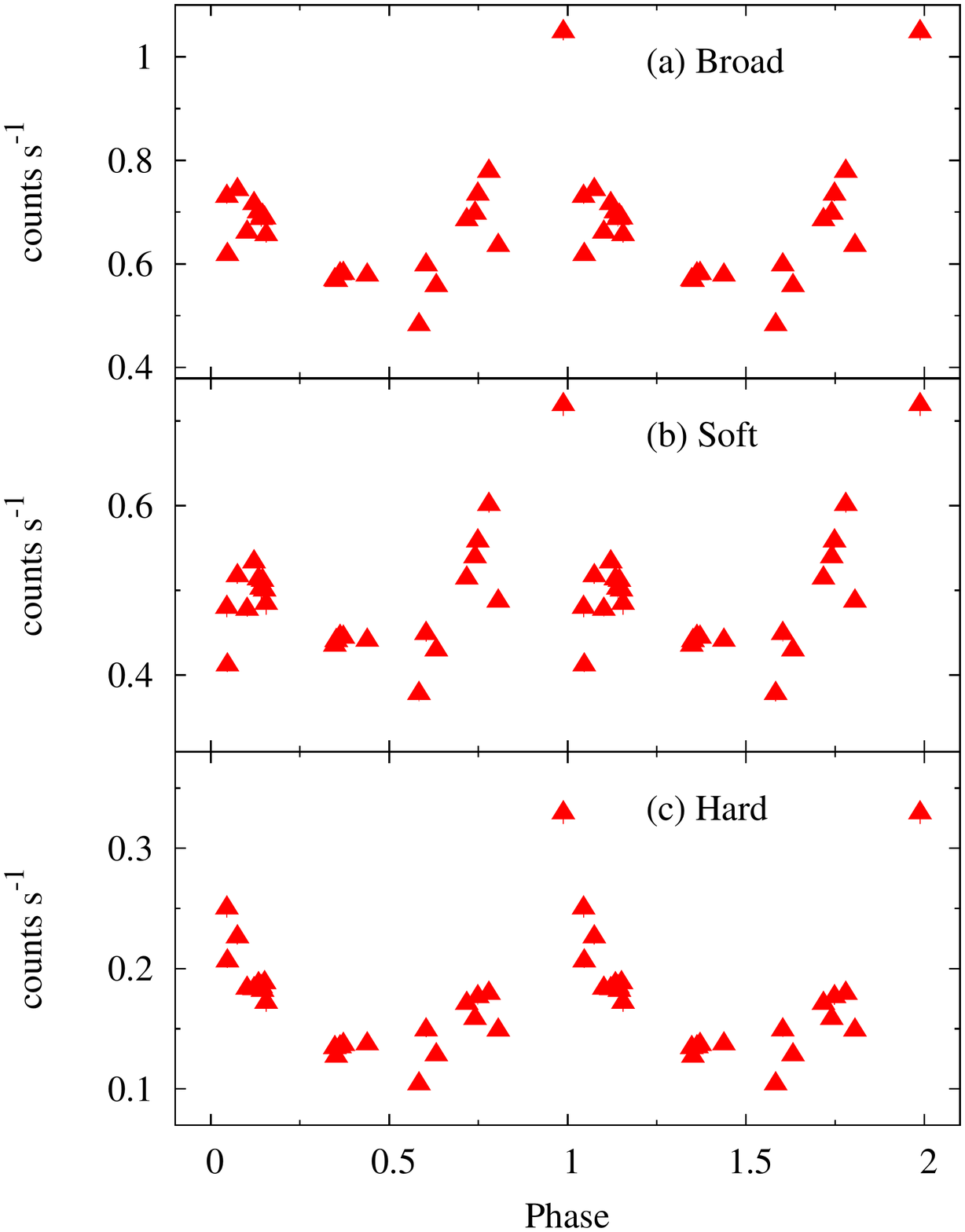}
        \caption{\textit{XMM-Newton}$-$MOS2}
        \label{fig:fig3c}
    \end{subfigure}
    \caption{Folded X-ray light curves in broad, soft, and hard energy bands as observed by (a) FI \textit{Suzaku} $-$ XIS, (b) \textit{Swift}$-$XRT, and (c) \textit{XMM-Newton}$-$MOS2.}
    \label{fig:fig3}
\end{figure*}

\begin{figure}
\includegraphics[width=\columnwidth,trim={0.0cm 2.8cm 0.0cm 4.7cm}]{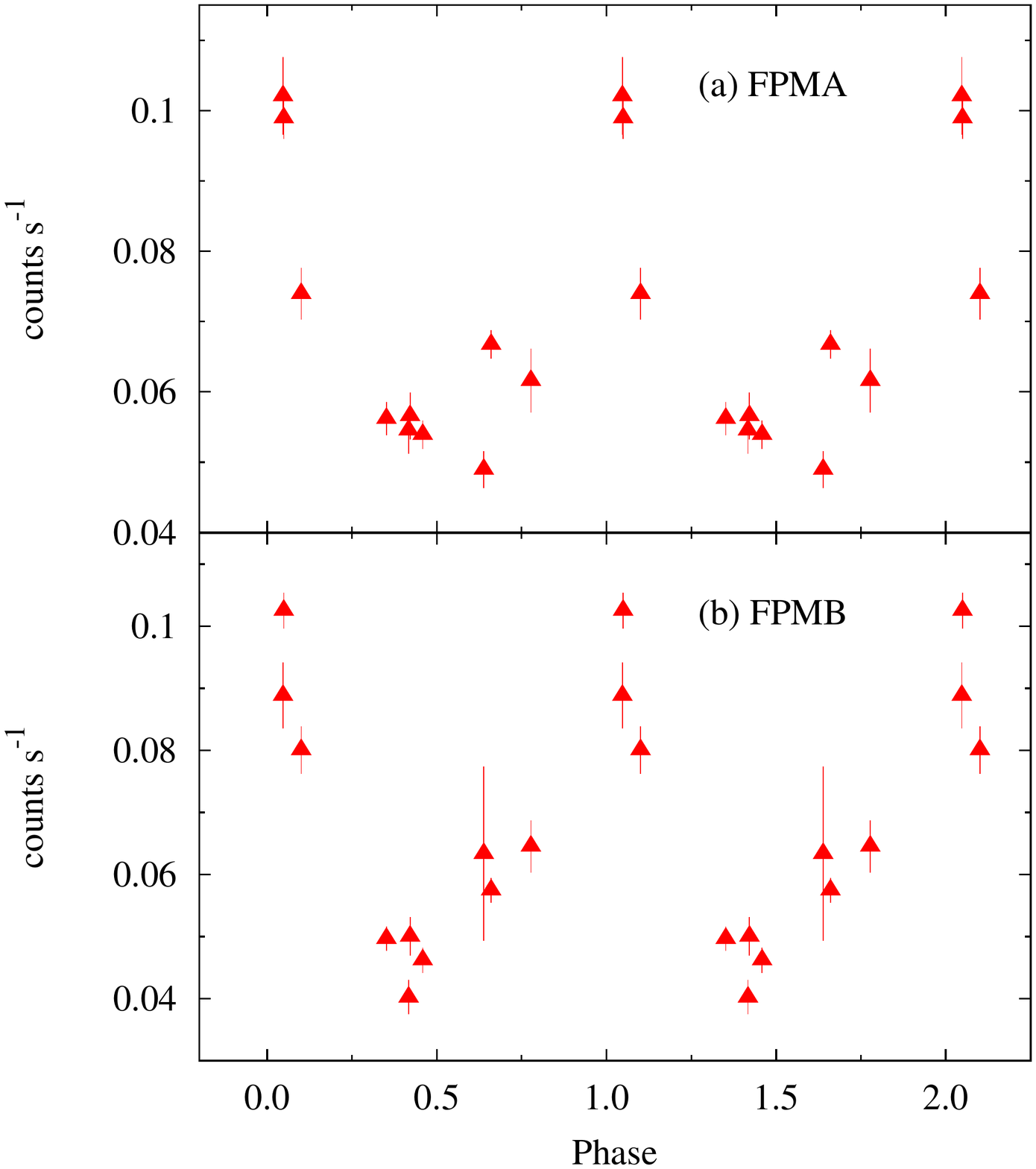}
\caption{Folded X-ray light curves in 3.0$-$10.0 keV energy range as observed by \textit{NuSTAR} $-$ FPMA (upper panel) and FPMB (lower panel). \label{fig:fig2}}
\end{figure}

The background subtracted light curves as observed by FPMA and FPMB onboard \textit{NuSTAR} were obtained in 3.0$-$10.0 keV and 10.0$-$78.0 keV energy bands. The median of the ratio of the count rate to  the corresponding error in 3.0$-$10.0 keV energy range was estimated to be 3.51 while it was 2.00 in 10.0$-$78.0 keV energy band. Therefore, we have considered \textit{NuSTAR} light curves only in 3.0$-$10.0 keV energy band for further analysis (see also Sect.~\ref{lookatspectra}). The background subtracted X-ray light curves were folded using the ephemeris HJD$=$2,451,598.0+207.85E given by \citet{Gamen2006}. The zero phase in the folded light curves corresponds to the time of the periastron passage. The X-ray light curves observed by FI \textit{Suzaku}$-$XIS, \textit{Swift}$-$XRT, and \textit{XMM-Newton}$-$MOS2 were folded in (a) broad, (b) soft, and (c) hard energy ranges and are shown in Figs.~\,\ref{fig:fig3a}, \ref{fig:fig3b}, and \ref{fig:fig3c}, respectively. Each point in the folded light curves corresponds to the average count rate of an observation ID. In Fig.~\ref{fig:fig3b}, the folded X-ray light curve of \textit{Swift}$-$XRT, red filled circles and blue triangles mark the observations made with a time gap of $\sim$7 years as in Fig.~\ref{fig:fig1}. It is evident that the count rates before and after this gap overlap fairly well, suggesting a reasonably stable behaviour of the phase-dependent X-ray emission over several orbits. Moreover, it also points towards the absence of any third component on a wider orbit, which would have been a potential cause of longer-term modulations (on top of that of the 208-d binary) in the X-ray light curve. Phase locked variations were seen in all the folded X-ray light curves and a similar kind of trend is observed. In the folded light curves of FI \textit{Suzaku}$-$XIS, \textit{Swift}$-$XRT, and \textit{XMM-Newton}$-$MOS2, initially the count rate increases around the periastron passage followed by a very sharp decrease in broad and soft energy bands just after the periastron and then it recovers to an average value around the apastron. On the other hand, the folded light curves as observed by \textit{NuSTAR}$-$FPMA and FPMB in 3.0$-$10.0 keV energy band are as shown in Fig.~\ref{fig:fig2}. Here again, the count rate decreases gradually while moving from periastron to apastron. The ratio of the maximum to the minimum count rate in broad, soft, and hard energy bands were found to be 2.48, 2.48, and 2.47 for FI \textit{Suzaku}$-$XIS, 4.46, 5.37, and 4.83 for \textit{Swift}$-$XRT, and 2.2, 1.9, and 3.2 for \textit{XMM-Newton}$-$MOS2 observations, respectively. However, for \textit{NuSTAR}$-$FPMA and FPMB observations in 3.0$-$10.0 keV energy range, this ratio is 2.1 and 2.5, respectively.


\subsection{Flux vs. binary separation}
\begin{figure*}
	\centering
	\begin{subfigure}[b]{0.32\textwidth}
		\captionsetup{skip=-31pt}
		\centering
		\includegraphics[width=6.4cm,trim={0cm -2.0cm 0.5cm 4.0cm}]{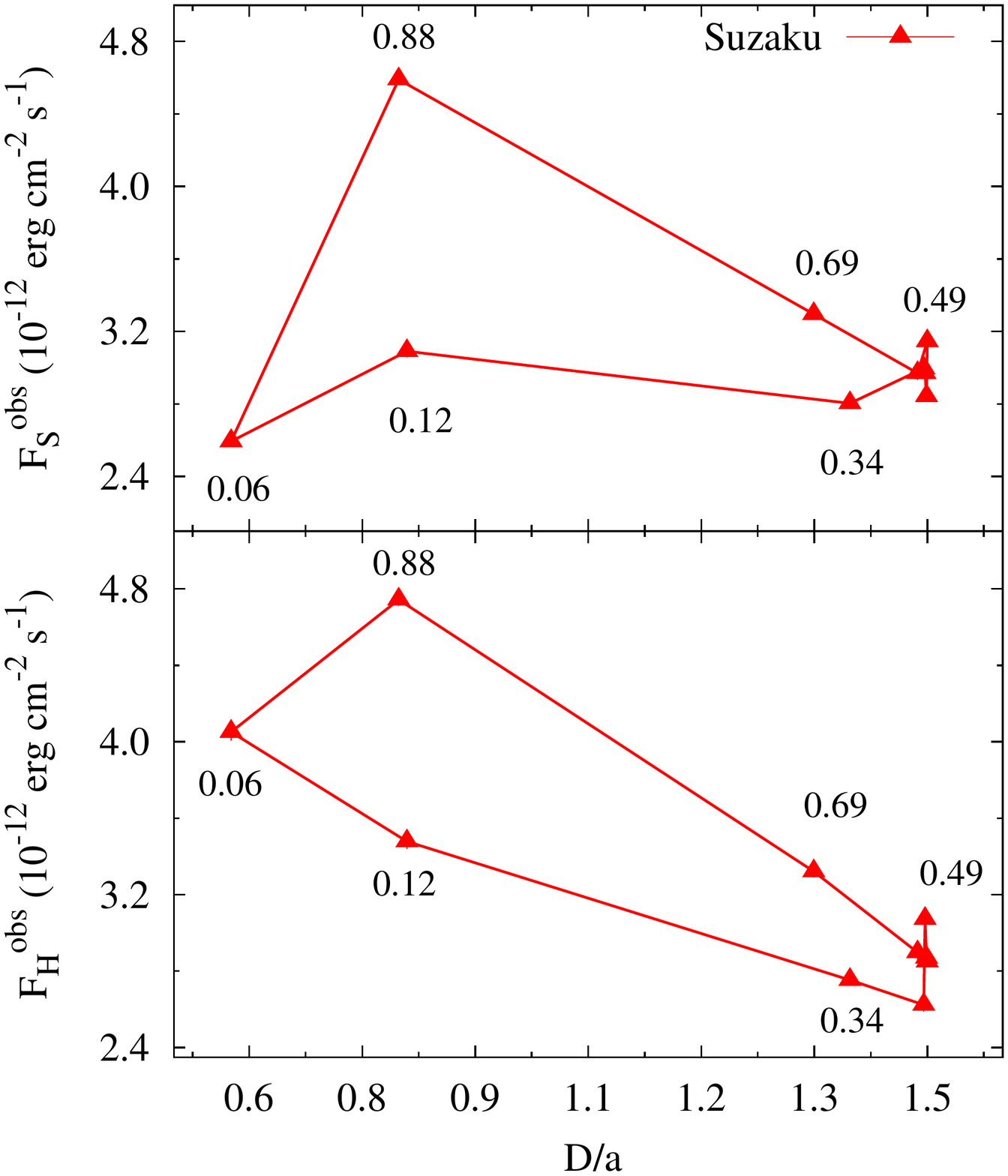}
		\caption{\textit{Suzaku}$-$XIS}
		\label{fig:fig8a}
	\end{subfigure}
	\hfill
	\begin{subfigure}[b]{0.32\textwidth}
		\captionsetup{skip=-31pt}
		\centering
		\includegraphics[width=6.4cm,trim={0cm -2.0cm 0.5cm 5.0cm}]{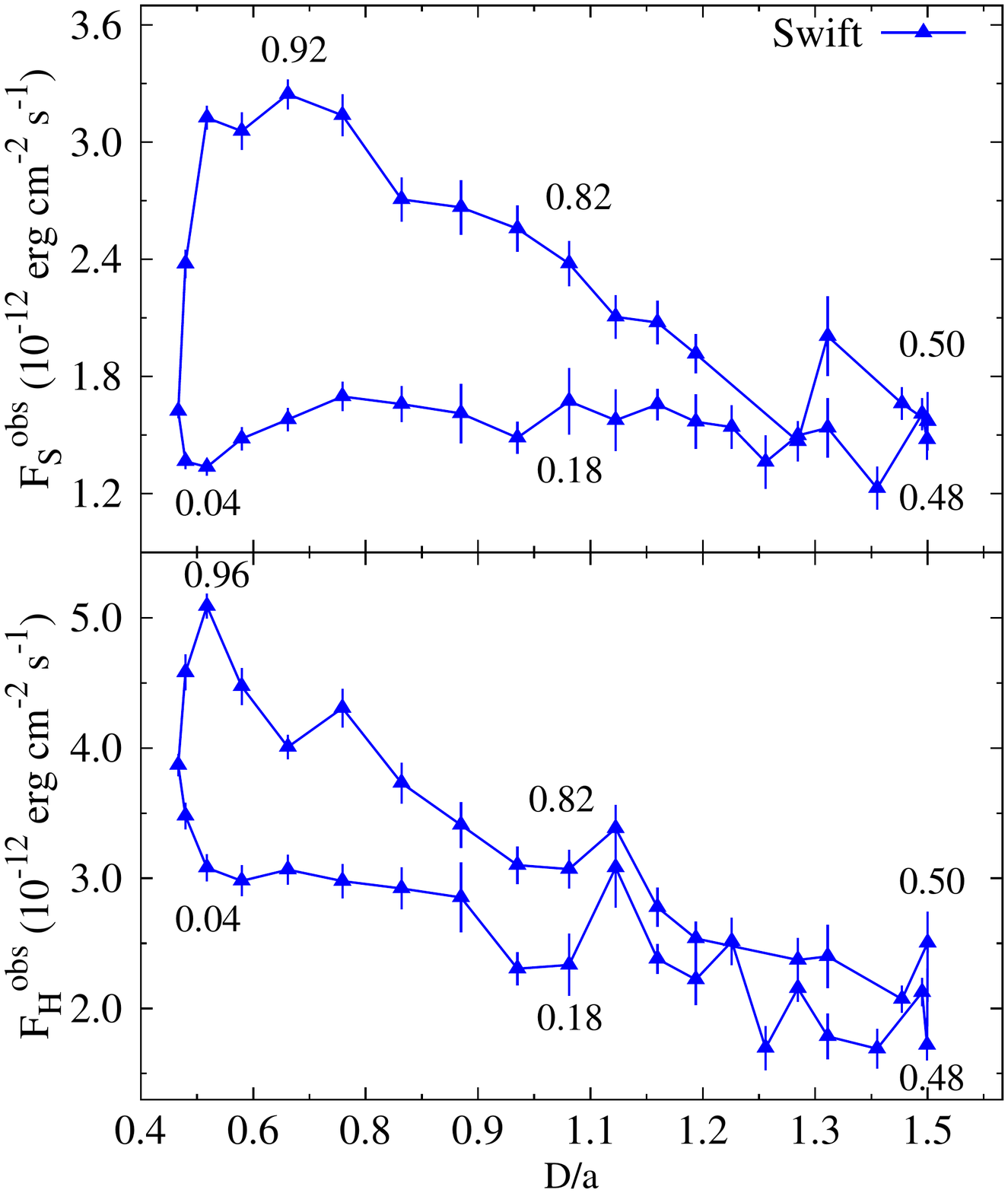}
		\caption{\textit{Swift}$-$XRT}
		\label{fig:fig8b}
	\end{subfigure}
	\hfill
	\begin{subfigure}[b]{0.32\textwidth}
		\captionsetup{skip=-31pt}
		\centering
		\includegraphics[width=6.4cm,trim={0cm -2.0cm 0.5cm 6.0cm}]{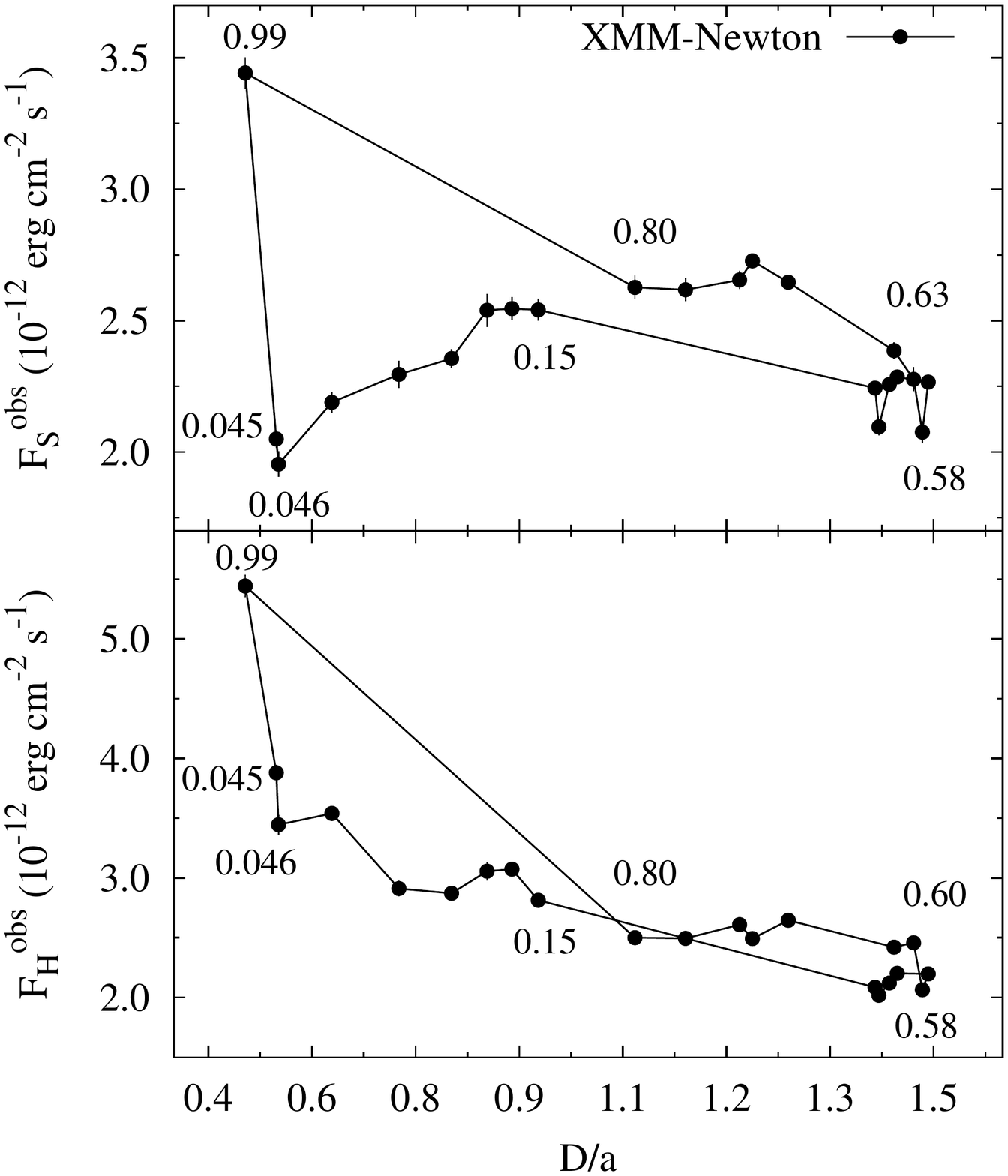}
		\caption{\textit{XMM-Newton}$-$EPIC}
		\label{fig:fig8c}
	\end{subfigure}
	
\vskip 15pt
	\begin{subfigure}[b]{0.32\textwidth}
		\captionsetup{skip=-10pt}
		\centering
		\includegraphics[width=6.4cm,trim={0.0cm +0.0cm 0.5cm 4.0cm}]{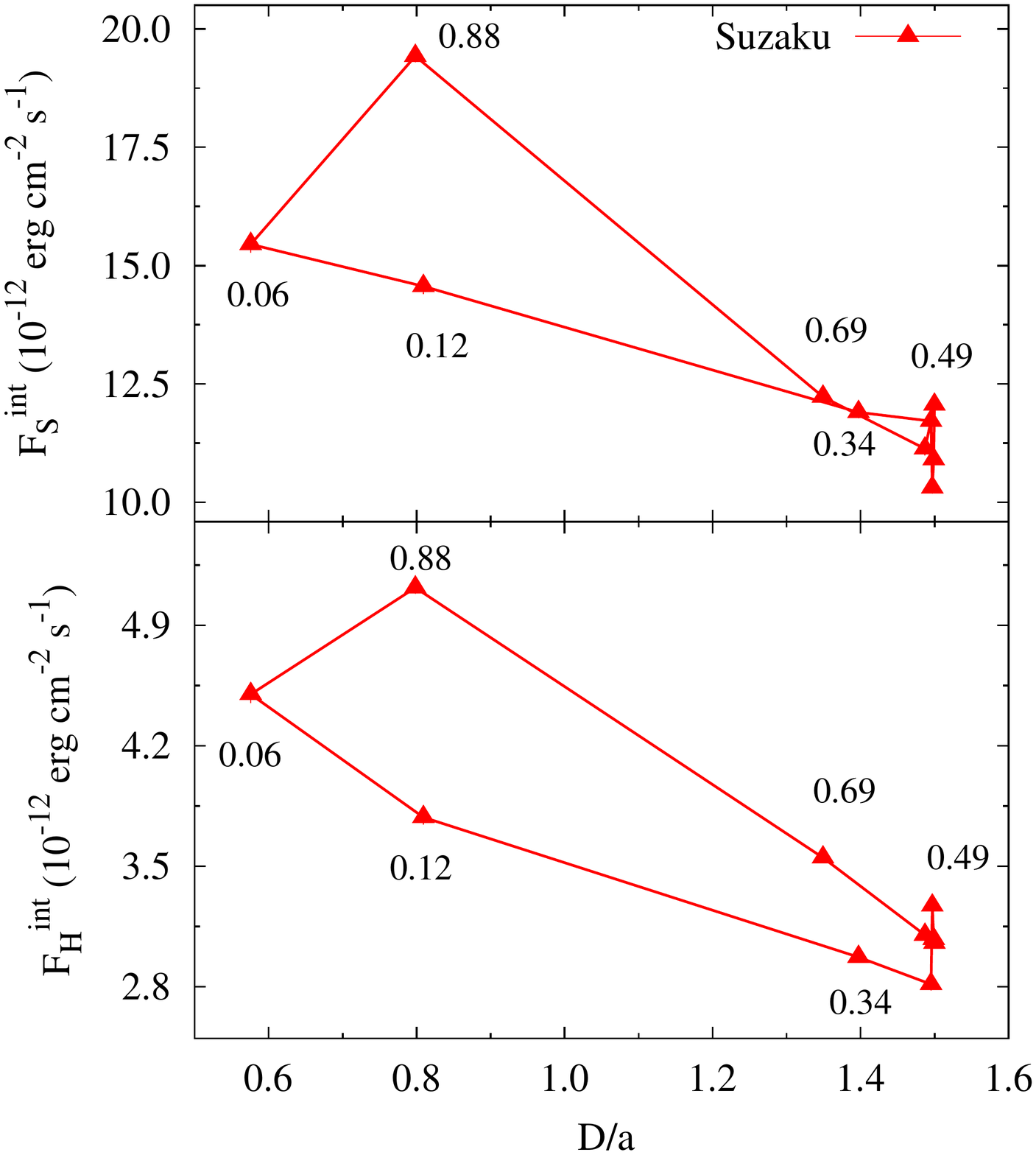}
		\caption{\textit{Suzaku}$-$XIS}
		\label{fig:fig8d}
	\end{subfigure}
	\hfill
	\begin{subfigure}[b]{0.32\textwidth}
		\captionsetup{skip=-10pt}
		\centering
		\includegraphics[width=6.4cm,trim={0cm +0.0cm 0.5cm 4.0cm}]{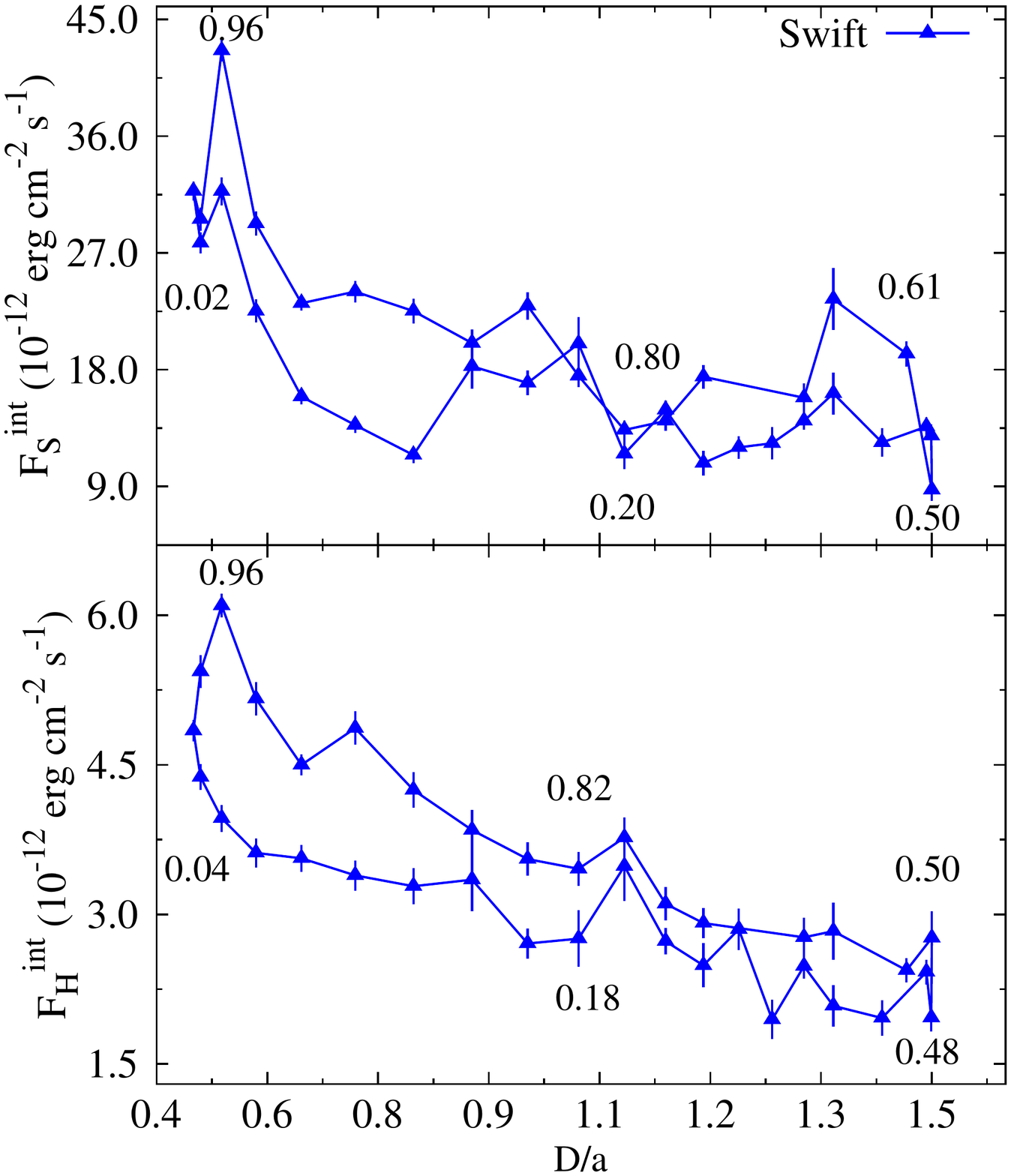}
		\caption{\textit{Swift}$-$XRT}
		\label{fig:fig8e}
	\end{subfigure}
	\hfill
	\begin{subfigure}[b]{0.32\textwidth}
		\captionsetup{skip=-10pt}
		\centering
		\includegraphics[width=6.4cm,trim={0cm 0.0cm 0.5cm 4.0cm}]{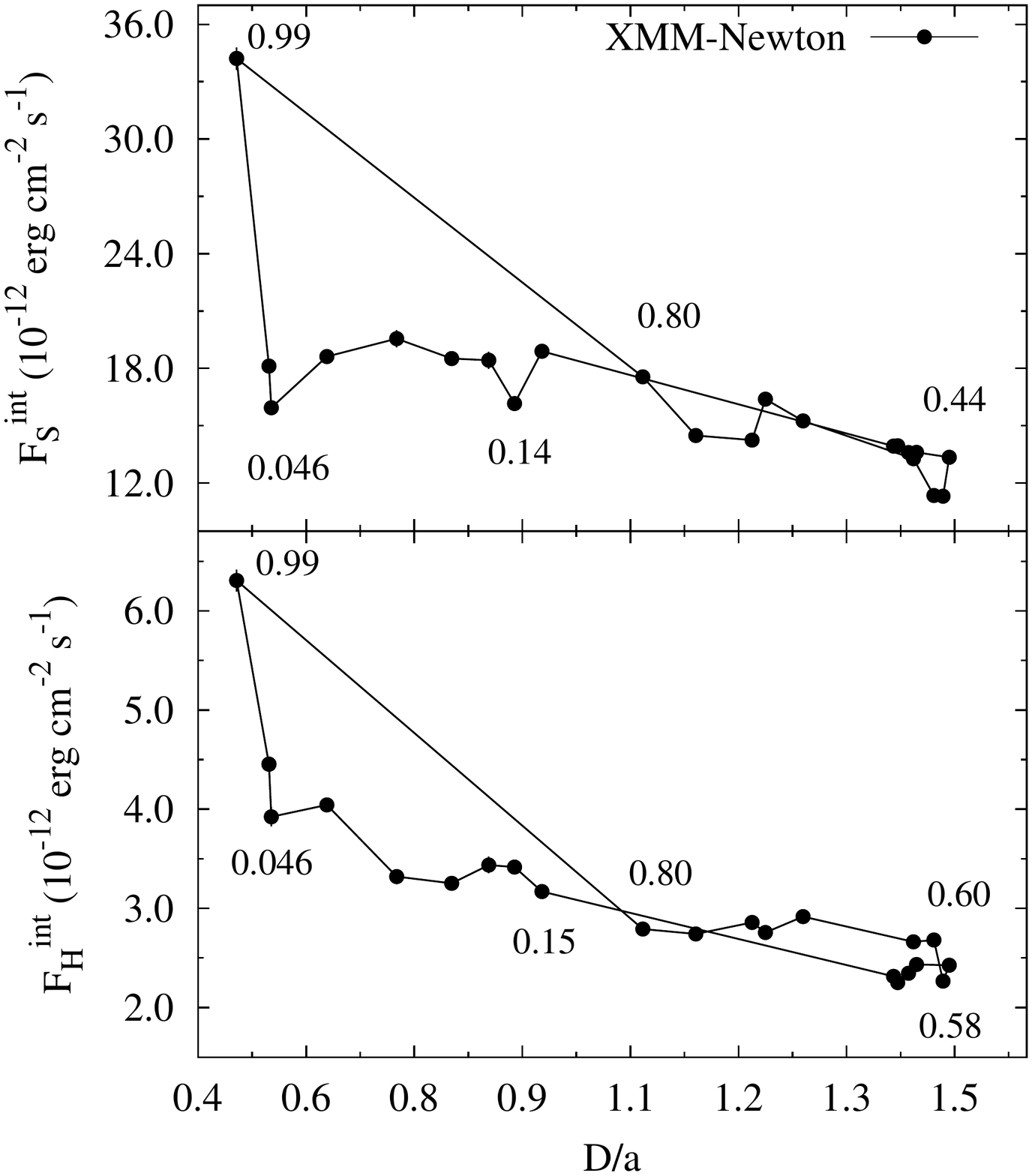}
		\caption{\textit{XMM-Newton}$-$EPIC}
		\label{fig:fig8f}
	\end{subfigure} 
	\caption{Observed (top row) and intrinsic (bottom row) X-ray flux of WR 25 as a function of binary separation in soft ($F_{S}$) and hard ($F_{H}$) energy bands as observed by \textit{Suzaku}$-$XIS, \textit{Swift}$-$XRT, and \textit{XMM-Newton}$-$EPIC. The orbital phases corresponding to a few data points have also been mentioned in the figures.}
	\label{fig:fig8}
\end{figure*}

The variation of observed and intrinsic X-ray fluxes in soft ($F_{S}$) and hard ($F_{H}$) energy bands with binary separation (D, i.e. the binary separation normalized to the semi-major axis a) as observed by \textit{Suzaku}, \textit{Swift}, and \textit{XMM-Newton} is shown in Fig.\,\ref{fig:fig8}. The orbital phases corresponding to a few observations are also mentioned in the figure. It is evident from these plots that the intensity of the X-ray emission varies as the two components of this binary system move towards and away from each other. These variations were found to be deviated from the linear trend. In order to find the significance of deviations from the linear trend a $\chi^2$-test was performed on the best fit straight line. We found that deviation was more than 99.9 per cent confidence level from linear trend for all the observations in both soft and hard energy bands. $F_{S}$ seems to be more affected as compared to $F_{H}$. The most pronounced variation in the emission occurs close to periastron passage, however, around the apastron, there is not much difference. \textit{Swift} observations has the dense coverage of the orbital period of WR 25 than other satellites' observations. Therefore, it will be more appropriate to discuss the  \textit{Swift}  observations. For \textit{Swift}, the maximum value of $F_{S}^{obs}$ occurred at phase 0.92 and dropped sharply to the phase 0.04. However, for $F_{H}^{obs}$, the maximum and minimum values were measured at phases 0.96 and 0.48 (Fig.\,\ref{fig:fig8b}), respectively. Both $F_{S}^{int}$ and $F_{H}^{int}$ were highest at phase 0.96 (Fig.\,\ref{fig:fig8e}) \textit{i.e.} before the periastron passage but these were lowest at phases 0.50 and 0.48, respectively. The plots for \textit{Suzaku} (Figs.\,\ref{fig:fig8a} and \ref{fig:fig8d}) and \textit{XMM-Newton} (Figs.\,\ref{fig:fig8c} and \ref{fig:fig8f}) also express a similar variation pattern but the poorer orbital sampling (especially close to periastron) prevents any accurate determination of the position of $F_{S}$ and $F_{H}$ extrema. We do not show the variation of $F_{S}$ and $F_{H}$ with D for all satellites data together since there are some differences between the individual satellite results, probably because of noise and cross-calibration effects, but a similar trend is followed by each.

\begin{figure*}
	\centering
	\begin{subfigure}[b]{0.55\textwidth}
		\centering
		\includegraphics[width=7.50cm,trim={3cm 1.8cm 7cm 16.5cm}]{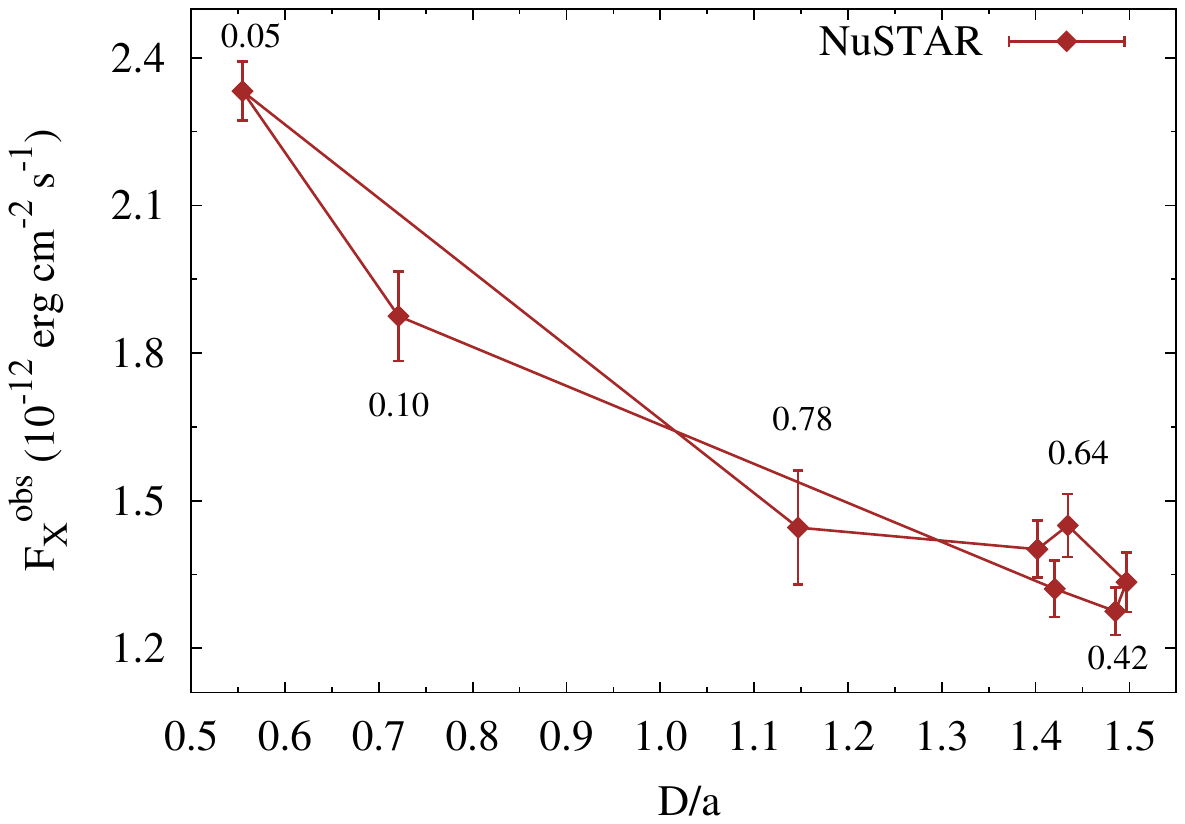}
		\caption{Observed X-ray flux variation}
		\label{fig:fig9a}
	\end{subfigure}
	\begin{subfigure}[b]{0.40\textwidth}
		\centering
		\includegraphics[width=7.50cm,trim={3cm 1.8cm 7cm 16.5cm}]{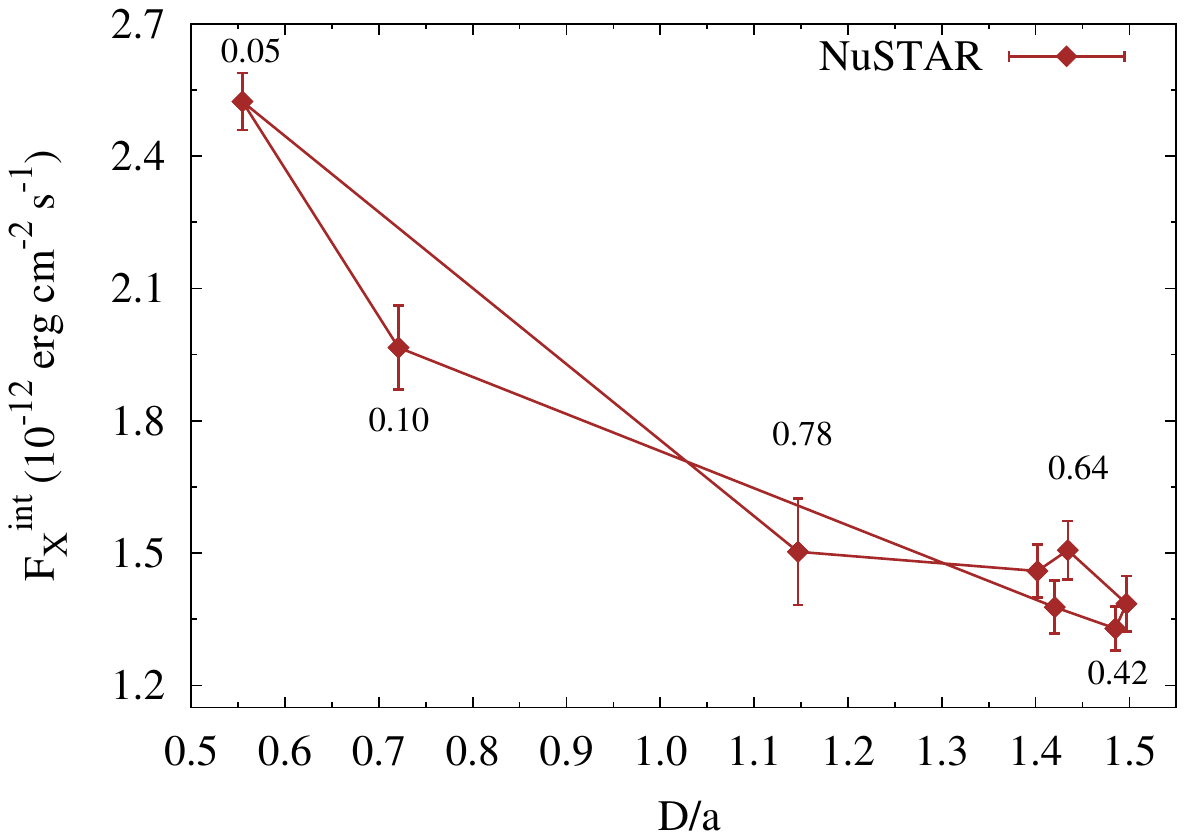}
		\caption{Intrinsic X-ray flux variation}
		\label{fig:fig9b}
	\end{subfigure}
	\caption{(a) Observed and (b) intrinsic X-ray flux as a function of binary separation in 3.0$-$10.0 keV ($F_{X}$) energy band as observed by \textit{NuSTAR}$-$FPMs. The orbital phases corresponding to a few data points have also been mentioned in the figures.}
	\label{fig:fig9}
\end{figure*}

The variation of $F_{X}^{obs}$ and $F_{X}^{int}$ as a function of normalized D as observed by \textit{NuSTAR} is shown Figs.\, \ref{fig:fig9a} and \ref{fig:fig9b}, respectively, and there is not much difference in both values at all the orbital phases. It appears that $F_{X}$ varied with D almost linearly and there is not any significant deviation in X-ray emission as two components of WR 25 move around the periastron. But a careful inspection of plots reveals that \textit{NuSTAR} observations did not cover much of the orbit around the periastron passage. After the phase 0.78, \textit{NuSTAR} observed WR 25 at phase 0.05 and then at phase 0.1. Therefore, the flux information in between the orbital phases 0.78 to 0.05 is missing in the part of the orbit where the most pronounced changes in the X-ray emission are expected.   

\section{Discussion}\label{disc}

\subsection{Variability of the thermal X-ray emission}
Our analysis, in general agreement with previous studies, emphasizes a few features deserving to be discussed and commented in their appropriate context.

The folded X-ray light curve of \textit{Swift}-XRT measurements (which offers the best orbital sampling) presents a peaked maximum close to the periastron passage, followed by a decrease of the emission on the way to apastron, suggesting an X-ray emission dependent on the stellar separation. This maximum is observed at orbital phases 0.943, 0.943, and 0.973, respectively in the broad, soft, and hard energy bands. The investigation of the separation-dependent variations must be based on quantities independent of the ISM as well as local absorption by the stellar wind (see below). A first indicator should be the X-ray flux measured in the hard band, weakly affected by photoelectric absorption. It is clear that the hardest emission is produced by the colliding-wind region and therefore should be affected by the varying separation, while the soft emission is produced by both the individual stellar winds and the wind-wind interaction region. The evolution of the hard X-ray flux as a function of distance is shown in the lower panels of Figs.\,\ref{fig:fig8a} to \ref{fig:fig8c}, and in Fig.\,\ref{fig:fig9a}. An alternative way would consist in plotting the intrinsic values (i.e. corrected for both local and interstellar absorptions) as a function of distance as shown in Figs.\,\ref{fig:fig8d} to \ref{fig:fig8f} and in Fig.\,\ref{fig:fig9b}. A third approach is to follow the evolution of the normalization constant corresponding to the hard component ($norm_2$, relevant for the colliding winds emission) as a function of distance. The general trends shown by these three indicators, especially when only hard X-ray fluxes are considered, are indeed very similar as expected.

Hydrodynamic models of wind interactions in massive binaries predict that WCR can behave either adiabatically or radiatively \citep{Stevens1992}. In shorter period systems, instabilities arise in the interaction region which leads to a quite turbulent wind collision and the X-ray luminosity then follows a relation of the form $L_{X} \propto f\dot{M}v^{2}$, where $f$ is the fraction of the wind involved in the collision, $\dot{M}$ is the mass loss rate, and $v$ is the pre-shock wind velocity. However, in the adiabatic regime, the interaction appears smoother and the X-ray luminosity scales as $L_{X} \propto \dot{M}^{2}v^{-3.2}D^{-1}$, where D is the binary separation \citep{Luo1990}. This behaviour is expected for binaries with orbital periods longer than few days. Since WR 25 has the orbital period of $\sim$ 208 days, therefore, it should follow the latter X-ray luminosity relation and hence the intrinsic flux is expected to be maximum at periastron, where the plasma density is higher. Based on the limited data set, \citet{Gosset2007} found that the hard X-ray flux of WR 25 increases linearly with the inverse of the relative binary separation using \textit{XMM-Newton} data, in full agreement with the expected behavior for an adiabatic case.

Light curves presented in Fig. \ref{fig:fig8} present a behaviour reminiscent of a hysteresis, even though this is not well supported by Fig. \ref{fig:fig9} characterized by a poorer orbital phase sampling. \citet{Pittard2010} predicted hysteresis behaviour in the variation of the X-ray luminosity from colliding winds as a function of the binary separation for eccentric systems. Though the model of \citet{Pittard2010} is designed for eccentric O+O binaries with period less than 10 days and weaker stellar winds, we find the similar features in the X-ray emission of WR 25. The origin of this effect is an asymmetry in the wind collision region about the line of centres, especially close to periastron passage. In the specific case of WR 25, the orbital velocity close to periastron reaches its maximum, while radiative affects (such as radiative inhibition and sudden radiative breaking, see below in this section) can reduce substantially the pre-shock wind velocity. The combination of these two effects is expected to result in an increase of the ratio between the orbital and wind velocities at periastron, which favors a distortion of the wind collision region. Such a distortion is expected to lead to a significant departure with respect to pure symmetry about the line of centers, causing the hysteresis. If the trends shown in Fig. \ref{fig:fig8} are real, one may thus tentatively transpose the case described by \citet{Pittard2010} to provide a likely interpretation context to our observations, provided we remember the basic idea behind this effect has to be slightly adapted to take into account some specificities of the system. The emission is brighter when the two components move towards each other than when they separate again. It may happen because of a higher pre-shock wind velocities when stars are getting closer, favoring a brighter hard X-ray component. As the two stars come close to each other, their winds collide with lower terminal velocities. The lower pre-shock velocity is less favorable to the hard X-ray emission that appears thus weaker. Therefore, the maximum X-ray luminosity is observed before the periastron passage. When the two components move away from each other after periastron, less hard X-ray emission is observed which reflects the smaller pre-shock wind velocities attained at earlier orbital phases, reducing the emission measure of the plasma that is warm enough to produce radiation significantly above 2\,keV. We have however to caution that this interpretation relies notably on a significant phase-locked variation of the pre-shock velocity, which is not fully supported by our measurements of the post-shock plasma temperature, as detailed in the discussion below in this section.

Our analysis shows that the X-ray emission deviates from the linear 1/D trend close to periastron, suggesting a departure from the adiabatic regime when the stellar separation is shorter. It is worth checking the evolution along the orbit of the cooling parameter ($\chi$) defined by \citet{Stevens1992} as the ratio of the cooling time of the post-shock gas to the typical escape time from the shock region. It is expressed as $ \chi = v^{4} D/\dot{M}$, where $v$ is the pre-shock wind velocity in 1000 km\,s$^{-1}$ units, $D$ is the distance from the star to the shock in $10^{7}$ km and $\dot{M}$ is the mass loss rate in 10$^{-7}$ M$_{\odot}$\,yr$^{-1}$ \citep{Stevens1992}. For $\chi\ll1$, the gas cools rapidly and the collision is considered to be radiative while for the adiabatic case, $\chi \geq 1$. The switch of $\chi$ down to radiative values can be further enhanced by radiative effects such as radiative inhibition and sudden radiative braking. These are related to the presence of the radiation field of the companion star and it has been shown that they may reduce the pre-shock velocity of colliding winds. Radiative inhibition involves the reduction in the initial acceleration of the stellar wind by the radiation field of its companion \citep{Stevens1994}. It is more suitable for close O\,+\,O binaries with comparably strong optically thin winds. However, sudden radiative braking is more favoured in WR + O binaries where the wind of the primary star is suddenly decelerated by the radiative momentum flux of its companion as it approaches the surface of that star \citep{Gayley1997}. It constitutes a more severe interaction that can significantly alter the bow shock geometry close to periastron passage. Since the cooling parameter is proportional to $v^{4}$, the lowering of the pre-shock velocity through this effect can have a significant impact on the shock regime, leading to a reinforcement of the radiative regime around periastron.

\begin{figure}
	\includegraphics[width=\columnwidth,trim={1.0cm 2.0cm 0.0cm 3.0cm}]{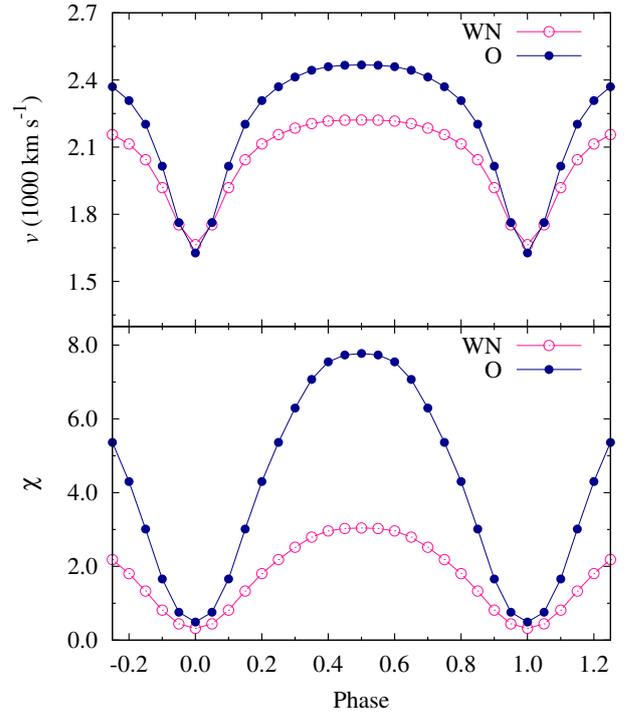}
	\caption{Expected evolution of the pre-shock wind velocity ($v$) and cooling parameter ($\chi$) with  the orbital phase for both  components of WR 25. Here, secondary component of WR 25 is an O4 supergiant.
		\label{fig:fig12}}
\end{figure}

We estimated $v$ and $\chi$ for both components of WR 25 at the position of the WCR by using the standard $\beta$-velocity law (thus without any influence by radiative effects mentioned in the previous paragraph). To achieve this, the typical values of the various stellar parameters for the WN6ha (primary) and O4 (secondary) stars were considered as given in Table\,\ref{stellar_par}. $\chi$ is expected to change as a function of the orbital phase because (i) the stellar separation changes and (ii) the pre-shock velocity is likely to change if the winds collide before they reach their terminal velocity. Our estimates show that within a 0.10 phase interval around periastron, both shocked winds should become slightly radiative with a $\chi$ value slightly lower than 1, especially if the O4 star is assumed to be a supergiant. $\chi$ remains higher than 1 during the remaining parts of the orbit for both winds. The evolution of $v$ and $\chi$ with the orbital phase for both components of WR 25 are shown in Fig. \ref{fig:fig12} assuming  secondary star as an O4 supergiant. Here, we see that  $v$ decreases by 25 per cent and 34 per cent for WN- and O- star, respectively, from apastron to periastron. One should however caution that the above discussion is valid for expectations based on simple principles, but one has to note a clear discrepancy between the expected trend of the pre-shock velocity illustrated in Fig. \ref{fig:fig12} and the measurements of the post-shock plasma temperature reported on in Fig.\,\ref{fig:fig11}. As the post-shock plasma temperature should scale with the square of the pre-shock velocity, a significant and smooth phase-locked evolution of the plasma temperature should be measured, but this is not observed. Provided the measured plasma temperature is an adequate proxy of the pre-shock velocity in the wind-wind interaction region, the simple interpretation context described here fails to explain all the emission properties described in this study. In order to achieve a more appropriate interpretation of the X-ray emission of WR\,25 and of its variability, the support of more sophisticated modelling is strongly required. In particular, the pre-shock velocity along the colliding-wind region appears as a critical physical quantity strongly influencing the overall X-ray emission of the system. An adequate dynamical modelling is needed to reconcile the apparent contradiction between the trends shown by Fig.  \ref{fig:fig12} and Fig. \ref{fig:fig11}.
\begin{table*}
	\caption{List of the stellar parameters adopted for the estimation of cooling parameter ($\chi$) at the WCR of WR\,25.	\label{stellar_par}}
	\begin{center}
		\begin{tabular}{l c c c c c}
			\hline
			\hline
			Parameter & WN6ha & \multicolumn{3}{c}{O4 (secondary)} & Reference \\
				&(primary)	&	Dwarf	& Giant	& Supergiant\\
			\hline
			\vspace*{-0.2cm}\\
			Mass ($M_\odot$) & - & 46.16 & 48.80 & 58.03 & 2 \\
			Radius ($R_\odot$) & 20 & 12.31 & 15.83 & 18.91 & 1,\,2 \\
			Mass-loss rate ($M_\odot$ yr$^{-1}$) & $10^{-5}$ & $10^{-5.836}$ & $10^{-5.540}$ & $10^{-5.387}$ & 1,\,2 \\
			Terminal wind velocity (km s$^{-1}$) & 2500 & 3599 & 2945 & 2877 & 1,\,2 \\
			$\beta$-velocity law & 1.00 & 0.90 & 0.90 & 0.92 & 1,\,2 \\
			\hline
		\end{tabular}
		\begin{tablenotes}
			\small
			\item \textbf{Notes:} The radius of the primary component was derived using the values of luminosity ($10^{6.18} L_{\odot}$) and temperature (45000 $K$) for WN6ha star from \citet{Crowther2007} assuming the black body emission.
			\item \textbf{References:} (1) \citet{Crowther2007}; (2) \citet{Muijres2012}
		\end{tablenotes}
	\end{center}
\end{table*}

Besides the trend discussed above related to the change of stellar separation, local absorption by the stellar winds material is also important to interpret X-ray light curves. The impact of this local absorption is clearly measured as a sudden and sharp drop in the X-ray flux that occurs shortly after periastron passage, in the broad and soft energy bands. According to the radial velocity curve of WR 25, presented by \citet{Gamen2006}, primary component of this binary remains towards the observer from phase $\sim 0.95$ to $0.27$. A sharp minimum is observed in $F_{S}^{ism}$ at orbital phase $0.04\pm0.01$ of WR 25. Therefore, this drop is the resultant of the strong local absorption faced when the WN wind is in front of the colliding-wind region, around this phase, from the observer's point of view. This feature was not so obvious in the results presented in the previous studies \citep{Gosset2007,Pandey2014} because of a sparser orbital sampling. This post-periastron minimum is not observed in the hard energy band, which is basically unaffected by photoelectric absorption.

We want to caution that the results obtained from the present study represent a very approximate picture of the actual physical condition of the system. In particular, one knows that the plasma temperature derived from our modelling approach consists of a rough average. Actually, the X-ray emitting region is made of the distribution of cells spreading over wide ranges of density and temperature. The post-shock temperature distribution is expected to peak along the line of centres where the pre-shock velocity of the colliding winds is the highest, with lower typical temperatures away from the apex of the shocks. Such a more realistic representation would be much too detailed considering the accuracy of the data used in this study. Our approach focuses thus on average and typical values, easily represented by the models and the modelling tools we used in this study. On the other hand, one also has to clarify that freezing the post-shock plasma temperature across the full spectral time series is at first sight at odd with the expectation of a varying pre-shock velocity in an eccentric system. In particular, the hardest component significantly accounting for the X-ray emission close to the line of centres should in principle be affected. However, the change in pre-shock velocity is not high enough to completely lead to dramatic changes in plasma temperatures, at least in a large part of the orbit and at the level of accuracy of the measurements allowed by the data (see Figure \ref{fig:fig11}). As a result, this approximation does not compromise the validity of our general interpretation. Concerning the element abundances, one should also keep in mind that the emitting plasma is coming from both stellar winds, and this is not straight forward to anticipate in which proportion, especially as a function of the orbital phase. In addition, the abundances of the absorbing plasma are also expected to vary depending on which stellar wind is in front. The required improvements in the spectral analysis to account for such effects are far beyond the information content of our data series and out of reach of our modelling tools. Our approach offers thus the advantage to provide a relevant description with a reasonably low number of free parameters, which constitutes a convenient requirement to perform the variability analysis described in this study.  

Finally, the long time basis of the data series investigated in this paper (about 16 years) suggests a fairly good consistency of the phase-folded X-ray emission even when observations distant by several years are considered. This indicates that the binary system scenario is fully satisfactory to explain the temporal behavior of WR 25, rejecting the idea that it might be a higher multiplicity system with an additional colliding-wind region (on a wider orbit) contributing to the overall thermal X-ray emission.

\subsection{Lack of non-thermal emission}\label{nonthermal}
It has been noticed that some colliding wind binaries also act as sources of particle acceleration in their wind collision region through the diffusive shock acceleration (DSA) mechanism which leads to the production of relativistic particles \citep{Debecker2013}. Relativistic electrons travelling in the magnetic field may give rise to synchrotron radio emission or they may also inverse comptonize the photospheric stellar light to X-rays or even soft $\gamma$-rays. This opens up the possibility that some non-thermal X-ray emission may be measured in colliding-wind binaries. As emphasized in this paper, and the same holds for other colliding-wind binaries, the soft X-ray emission (below 10\,keV) is dominated by thermal emission from the wind-wind interaction region. As a result, attempts to measure such a non-thermal X-ray emission should focus on hard X-rays, above 10\,keV \citep[see e.g.][]{Debecker2007}. However, the lack of significant X-ray emission revealed by \textit{NuSTAR} between 10 and 78\,keV provides evidence that no inverse Compton scattering emission is produced by WR 25 above the background level.

The availability of hard X-ray data allows however to derive upper limits on the count rate of the putative non-thermal X-ray emission between 10 and 78 keV for FPMA and FPMB instruments. We applied the procedure applied by \citet{DeBecker2014} to \textit{NuSTAR} datasets obtained at two extreme orbital phases, respectively close to periastron ($\phi$ = 0.049) and apastron ($\phi$ = 0.458). We filtered event lists using a circular spatial filter with  a radius of 30$''$ to measure the associated number of counts ($C$). This radius was selected because it corresponds to the half the Half Power Diameter (HPD) given by \citet{Harrison2013}. This extraction region is large enough to collect a count number still significant for Poisson statistics, and it is small enough to avoid any significant contamination by adjacent imaging resolution elements. This count number was then corrected for the encircled energy fraction corresponding to the extraction radius (50 per cent,  as we adopted an extraction radius corresponding to that fraction), and it was further corrected for the position dependent effective area of the FPMs. For the latter correction, we considered a median energy, i.e. 40 keV. According to \cite{Harrison2013}, at a distance of about 5.5$'$ from the on-axis position (see the observation log in Table\,\ref{log}) the effective area is about 45 per cent of its maximum (on-axis) value. We thus obtain a corrected count number ($C_{cor}$).  We then determined a count threshold ($C_{max}$) corresponding to a logarithmic likelihood ($L$) of 12, translating into a probability ($P$) to find a count number in excess of $C_{max}$ of about 6\,$\times$\,10$^{-6}$ ($L = -\ln\,P$), under the null hypothesis of pure background Poisson fluctuations. This criterion is frequently adopted as a threshold for statistical fluctuations. In practice, we iteratively estimated the logarithmic likelihood assuming Poisson statistics on the basis of the corrected count number and adopting a first guess for $C_{max}$. At every iteration, $C_{max}$ was adapted to converge to a logarithmic likelihood of 12, for a fixed $C_{cor}$ value. The difference between these two quantities ($C_{max}$ at $L = 12$ and $C_{cor}$) gives the maximum expected count excess. The division of the latter quantity by the effective exposure time gives the count rate ($CR$) upper limit on the putative emission, as shown in Table\,\ref{upperlimits}. This approach has the benefit to estimate a count excess on a statistically relevant basis.

The upper limits on the count rate for a hypothetical inverse Compton scattering X-ray emission can be converted into physical flux units assuming an adequate emission model. The non-thermal spectrum should be a power law, slightly affected by interstellar and local absorption in the softer energy band. Even though our upper limits on the count rate relies on the 10$-$78\,keV range, the IC process should contribute over the full X-ray range, and we estimated the flux between 0.1 and 100 keV. We used the WebPIMMS on-line tool\footnote{https://heasarc.gsfc.nasa.gov/cgi-bin/Tools/w3pimms/w3pimms.pl} assuming two different values of the photon index: 1.5 (the standard value for DSA in strong shocks, e.g. \citealt{Debecker2007}) and 1.7 (valid for the likely leptonic component in Eta\,Car's soft $\gamma$-ray spectrum, \citealt{Farnier2011}). Depending on the orbital phase, on the instrument and on the assumed photon index, we obtain intrinsic non-thermal X-ray fluxes in the range 7-9\,$\times$\,10$^{-13}$\,erg\,cm$^{-2}$\,s$^{-1}$. For a distance of 2.1\,kpc \citep[Gaia DR2;][]{Lindegren2018}, this converts into upper limits on the intrinsic luminosities in the range 3.6-4.7\,$\times$\,10$^{32}$\,erg\,s$^{-1}$.

These numbers deserve to be discussed in the context of the energy budget of colliding-wind massive binaries, and of their participation in non-thermal processes. The IC process (over the full X-ray domain) and the thermal X-ray emission (measured below 10 keV) share the same energy reservoir, i.e. the fraction of the wind kinetic power that is injected in the colliding winds. On the basis of previous measurements on colliding-wind binaries, the fraction of that energy that is converted into thermal X-rays  emerging from the winds is of the order of 1-10 per cent. A similar fraction is expected to apply for the energy injection into non-thermal particles, on the basis of theoretical considerations and of a comparison with the case of supernova remnants which share the same physics. However, IC scattering comes only from the energy injected into electrons, which constitutes only a fraction of the energy of the non-thermal particles, i.e. likely a few per cent of the energy injected in relativistic particles \citep[see][for a discussion of the energy budget of Particle-Accelerating Colliding-Wind Binaries]{Debecker2013}. Considering the high radiative energy density in the particle acceleration region, the energy injected in relativistic electrons should be dominantly radiated through IC scattering. The amount of energy injected into the IC process should thus, roughly, be at most equivalent to a few per cent of the energy radiated in thermal X-rays and escaping from the winds. However, our upper limits on the non-thermal X-ray emission are only a few per cent of the intrinsic thermal X-ray fluxes reported in this paper, and about a factor 10 lower than the thermal X-ray fluxes corrected for ISM absorption only (see Table\,\ref{tab:spectra}). These upper limits do not therefore provide better constraints than the anticipated educated guesses proposed by previous studies. Consequently, the limitations due to the sensitivity of \textit{NuSTAR} lead to quite loose upper limits, which do not provide stringent constraints on the non-thermal physics. Only systems with higher kinetic power, hence with a more abundant mechanical energy reservoir, would have a chance to be detected as non-thermal emitters in hard X-rays with \textit{NuSTAR}. As a result, present day non-detection constitute by no means a severe drawback for the scenario of non-thermal high energy emission from massive binaries.

\begin{table}
	\caption{Estimates of the upper limits on the count rate for \textit{NuSTAR} instruments at the position of WR\,25.\label{upperlimits}}
	\begin{center}
		\begin{tabular}{l c c}
			\hline
			\hline
			& FPMA & FPMB  \\
			\hline
			& \multicolumn{2}{c}{$\phi$ = 0.049}\\
			\vspace*{-0.2cm}\\
			$C$ (cnt)& 189 & 194 \\
			$C_{cor}$ (cnt) & 839 & 861 \\
			$C_{max}$ (cnt) & 969 & 993 \\
			$C_{max} - C_{cor}$ (cnt) & 130 & 132 \\
			Eff. exp. time (s) & 79400 & 79400 \\
			$CR$ (cnt\,s$^{-1}$) & 1.6\,$\times$\,10$^{-3}$ & 1.6\,$\times$\,10$^{-3}$ \\
			\vspace*{-0.2cm}\\
			\hline
			& \multicolumn{2}{c}{$\phi$ = 0.458}\\
			\vspace*{-0.2cm}\\
			$C$ (cnt)& 144 & 126 \\
			$C_{cor}$ (cnt) & 639 & 559 \\
			$C_{max}$ (cnt) & 753 & 666 \\
			$C_{max} - C_{cor}$ (cnt) & 114 & 107 \\
			Eff. exp. time (s) & 54500 & 54500 \\
			$CR$ (cnt\,s$^{-1}$) & 2.0\,$\times$\,10$^{-3}$ & 2.0\,$\times$\,10$^{-3}$ \\
			\vspace*{-0.2cm}\\
			\hline
		\end{tabular}
	\end{center}
\end{table}

On the other hand, the most active indicator for particle acceleration in massive binaries is synchrotron radio emission. Previous radio continuum observations of Southern massive stars allowed to measure a flux density of 0.90$\pm$0.15 mJy at a wavelength of 3 cm for WR 25 \citep{Leitherer1995,Chapman1999}. Only upper limits to the flux density could be obtained at longer wavelengths. No formal determination of the nature (thermal or non-thermal) of the radio emission from WR 25 could be achieved as it was detected at only one frequency, preventing therefore any spectral index determination \citep{Leitherer1997}. However, by combining all the measurements, \citet{Dougherty2000} gave a lower limit of $\alpha=-1.26$ to the radio spectral index of WR 25 (for a flux density dependence on the frequency defined as $S_\nu \propto \nu^\alpha$). \citet{Leitherer1995} argued that the observed 3 cm emission from WR 25 is of thermal nature by comparing the mass loss rate of WR 25 using models given by \citet{Wright1975} and \citet{Panagia1975} to those obtained by some independent techniques which assumed thermal nature of the emission. We calculated the expected thermal free-free radio emission from the WN wind following the \citet{Wright1975} approach, adopting the stellar parameters for a WN star wind quoted in Table\,\ref{stellar_par}. We also adopted an electron temperature equal to 50 per cent of the effective temperature \citep{drew1990}, for an effective temperature of 45000\,K (for a WN6h classification, \citealt{Crowther2007}). Our estimate of the flux density was corrected to take into account wind clumping, following the same approach as \citet{Debecker2018}. As noted by \citet{Puls2008}, mass loss determinations on the basis of radio flux density measurements should be reduced by a factor $\sqrt{f}$ to account for clumping  (where $f$ is the clumping factor). Accordingly, a clumped stellar wind characterized by a given mass loss rate will generate thermal radio emission with a flux density a factor $f^{2/3}$ greater than for a smooth, un-clumped configuration. We assumed a clumping factor of 4, valid for the outer parts of the wind where the thermal radio emission is produced \citep{Runacres2002}. The same procedure was followed to estimate the contribution from the O-star wind, using the wind parameters given in Table\,\ref{stellar_par} and an effective temperature of about 40000\,K \citep{Muijres2012}. As a result, we estimate that the observed cumulative flux density at 3 cm for both components winds, at a distance of 2.1 kpc, should be of the order of 0.60 mJy (about 0.50 and 0.10 mJy for the WN and the O components, respectively), in fair agreement (within uncertainties) with the measurement at that wavelength. This provides some significant support to the idea that the radio measurement of WR 25 is more likely made of thermal emission only, without the need to call upon any additional non-thermal contribution to interpret the measurements. The lack of non-thermal radio emission associated to WR 25 may be attributed either to an inefficient acceleration process or to a strong free-free absorption by the WN (and to some extent O) stellar wind material. The latter process constitutes indeed a very likely turn-over process for synchrotron spectra produced by massive binaries \citep[see e.g.][]{DeBecker2017}. At this stage, no hint for particle acceleration has been revealed for WR 25.

\subsection{Comparison with other systems}

CygOB2\#9 is another wind interacting source (O5$-$5.5I$+$O3$-$4III, e=0.71, $P_{orb}$=858.4 d) which shows a very clear 1/D variation of the X-ray flux as expected for very long period binary systems \citep{Naze2012}. A small deviation with respect to the 1/D trend  was observed for the hard X-ray flux at periastron. \citet{Naze2012} suggested that this might be a consequence of the collision becoming slightly radiative around the periastron due to increased wind density. They also found that the emission was somewhat softer near periastron and suggested that radiative inhibition and/or braking became efficient when the two binary components were closest. \citet{Parkin2014} confirmed this assertion for CygOB2\#9 and showed that wind acceleration is inhibited at all phases by the radiation field of the companion star. 9 Sgr (O3.5 V((f*))$+$O5-5.5 V((f)), e=0.71, $P_{orb}$=9.1 yr, \citealt{RauwBlomme2016}) is another long period colliding wind binary with a significant deviation from the expected 1/D behaviour close to periastron. 
Among shorter period systems, some massive binaries display a clear hysteresis effect in the dependence of the X-ray emission as a function of stellar separation \citep{Pittard2010}. These systems include CygOB2\#8a (O6If$+$O5.5III(f), e=0.21, $P_{orb}$=21.9 d, \citealt{DeBecker2006,Cazorla2014}); HD 152248 (O7.5(f)III$+$O7(f)III, e=0.13, $P_{orb}$=5.816 d, \citealt{Sana2004,Rauw2016}); HD 152218 (O9IV$+$O9.7V, e=0.26, $P_{orb}$=5.60 d, \citealt{Sana2008,Rauw2016}); WR 21a (WN5h$+$O3V, e=0.69, $P_{orb}$=31.680 d, \citealt{Gosset2016}); HD 166734 (O7.5If$+$O9I(f), e=0.618, $P_{orb}$=34.53 d, \citealt{Naze2017}). All of these systems, with shorter orbital periods, show larger X-ray flux values around apastron than periastron with a maximum value in between apastron and periastron. But in the case of CygOB2\#9 and WR 25, lower X-ray fluxes are observed at apastron than periastron with a maximum close to periastron only. This is expected as shorter period systems undergo stronger deviations from the expected 1/D behaviour as compared to longer period binary systems: on the one hand shorter period systems are more prone to be radiative, and on the other hand the dynamics of the shocks is more sensitive to the distortion produced by the higher orbital velocity in short period binaries \citep{Pittard2010}. 

From the point of view of non-thermal X-ray emission, our results are in line with previous non-detection already reported for massive binaries in the Cygnus region on the basis of \textit{INTEGRAL} observations \citep{DeBecker2007b}. However, our upper limits are more constraining than those derived with \textit{INTEGRAL} by about two orders of magnitude, thanks to the better sensitivity of \textit{NuSTAR}. The only massive binary system with a reported non-thermal high energy emission is Eta\,Car, with a hard X-ray luminosity of about 7\,$\times$\,10$^{33}$\,erg\,s$^{-1}$ \citep[][on the basis of \textit{INTEGRAL} observations]{Leyder2008}, thus significantly brighter than the upper limit we derived for WR 25. More recently, \citet{Hamaguchi2018} reported even on phase-locked non-thermal hard X-ray emission from Eta\,Car using \textit{NuSTAR} data. This lends support to the idea already mentioned in Sect.\,\ref{nonthermal} that only systems with a significantly high wind kinetic power (such as Eta\,Car) may reveal their non-thermal high energy emission, given the sensitivity of current instruments.

\section{Conclusions}\label{concl}
We carried out a deep X-ray study of WR 25 using the archival X-ray data from \textit{Suzaku}, \textit{Swift}, \textit{XMM-Newton}, and \textit{NuSTAR}. The time basis of the observations we investigated is more than 16 years, thus much longer than previous studies on this object.

In the soft X-ray domain below 10 keV, WR 25 is an over-luminous X-ray source as a result of the colliding stellar winds of the two components of the binary system. The system is brighter before periastron passage and becomes fainter when the line of sight passes through the dense wind of the WR star in front. Since harder X-rays are less affected by the enhanced column density, this atmospheric eclipse-like effect is not observed above 2\,keV. The analysis of the present data shows the wind collision is more or less adiabatic but with a significant deviation with respect to the expected 1/D dependence of the X-ray luminosity around periastron passage. This deviation may tentatively be explained by a lower pre-shock velocity close to periastron favoring a brief switch to the radiative regime, especially if the velocity drop is enhanced by sudden radiative braking. However, this interpretation is not supported by our measurements of post-shock plasma temperatures as a function of the orbital phase, which do not present any measurable drop close to periastron. Our long-term monitoring suggests the thermal X-ray emission overlaps fairly well when orbits distant in time are compared. This indicates that WR 25 is unlikely a triple system where an additional colliding-wind region in a wider orbit may contribute to the overall thermal X-ray emission.

Above 10 keV, \textit{NuSTAR} data do not reveal any X-ray emission attributable to a non-thermal high energy component due to inverse Compton scattering. The upper limits we derived on the putative non-thermal X-ray flux are in the range of 7-9\,$\times$\,10$^{-13}$\,erg\,cm$^{-2}$\,s$^{-1}$ (between 0.1 and 100.0\,keV), which is about a factor 10 lower than the ISM corrected thermal X-ray flux we measure below 10.0 keV. Considering one may expect the non-thermal energy injection into relativistic electrons (responsible for IC scattering) is at most a few per cent of the energy radiated in thermal X-rays, the presently accessible upper limits provide only loose constraints on the non-thermal high energy emission. A measurable IC emission could, however, be envisaged from systems characterized by quite large wind kinetic power, as in the case of Eta\,Car for instance. Non-thermal energy budget consideration for colliding-wind binaries show that even a significant energy injection would lead to a putative non-thermal emission below the sensitivity of \textit{NuSTAR} for most known systems. A sensitivity improvement of at least one order of magnitude is needed to access more stringent limits on the IC emission, or even have a chance to detect it for massive star systems with the most powerful winds.

\section*{Acknowledgements}
The present research work has used data obtained from the \textit{Suzaku} satellite, a collaborative mission between the space agencies of Japan (JAXA) and the USA (NASA), as well as from the \textit{NuSTAR} mission, a project led by the California Institute of Technology, managed by the Jet Propulsion Laboratory and funded by the National Aeronautics and Space Administration. These results are also based on the observations obtained with \textit{XMM-Newton}, an ESA science mission with instruments and contributions directly funded by ESA Member States and NASA. We acknowledge the use of public data from the \textit{Swift} data archive. The authors want to thank Dr. Eric Gosset for helpful discussions. We thank the anonymous referee for reading our paper and his/her useful comments.






\bsp	
\label{lastpage}

\end{document}